\documentclass[graybox, nosecnum]{svmult}

\usepackage{mathptmx}       
\usepackage{helvet}         
\usepackage{courier}        
\usepackage{type1cm}        
\usepackage{makeidx}         
\usepackage{graphicx}        
\usepackage{multicol}        
\usepackage[bottom]{footmisc}
\usepackage{hyperref}        
\usepackage{soul}            
\hypersetup{colorlinks=true,urlcolor=blue}
\usepackage[square,numbers]{natbib}
\makeindex             

\begin{document}
\title*{The AstroSat Observatory}
\author{Kulinder Pal Singh\thanks{corresponding author}}
\institute{Kulinder Pal Singh \at Tata Institute of Fundamental Reesearch, Mumbai 400005 and Indian Institute of Science Education and Research Mohali, SAS Nagar 140306, India., \email{kulinderpal@gmail.com}}

%
%
\maketitle
\abstract{AstroSat is India's first Ultra-violet (UV) and X-ray astronomy observatory in space.  The satellite was launched by the Indian Space Research Organisation on a Polar Satellite Launch Vehicle on 28 September 2015 from Sriharikota Range north of Chennai on the eastern coast of India.  AstroSat carries five scientific instruments and one auxiliary instrument.  Four of these consist of co-aligned telescopes and detectors mounted on a common deck of the satellite to observe stars and galaxies simultaneously in the near- and far-UV wavelengths and a broad range of X-ray energies (0.3 -- 80 keV). The fifth instrument consists of three X-ray detectors and is mounted on a rotating platform on a side that is oriented 90 degrees with respect to the other instruments to scan the sky for X-ray transients. An auxiliary instrument monitors the charged particle environment in the path of the satellite.}
\section{Keywords} 
Detectors, Telescopes, X-ray Astronomy – soft X-rays, hard X-rays,  X-ray Polarization, UV Astronomy - Far Ultra-Violet, Near Ultra-violet.
\section{Introduction}
\begin{figure}[h]
\centering
\includegraphics[width=120mm]{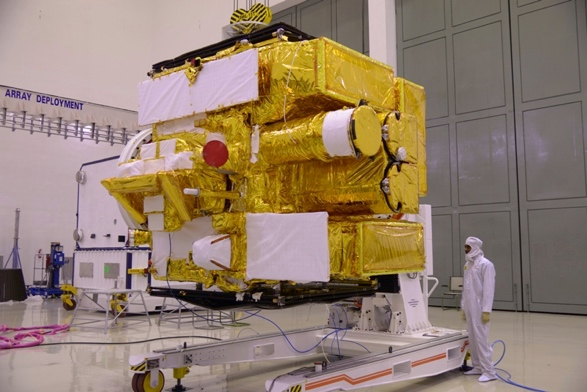}
\caption{The AstroSat spacecraft before launch (Credits: ISRO).}
\label{fig:spacecraft}       
\end{figure}

AstroSat was launched into an orbit with an inclination of 6 degrees and an altitude of 650 km on 28 September 2015 by the Polar Satellite Launch Vehicle, PSLV-XL C30.  The scientific instruments carried by the AstroSat consist of two Ultra-violet Imaging Telescopes (UVIT), three Large Area X-ray Proportional Counters (LAXPC), a Soft X-ray focusing Telescope (SXT), a Cadmium–Zinc–Telluride Imager (CZTI), a Scanning Sky Monitor (SSM), and an auxiliary detector called the Charged Particle Monitor (CPM).  The scientific payloads have a mass of 855 Kg and the total launch weight of AstroSat was 1513 kg.  An image of AstroSat with its payload (scientific and auxiliary instruments) in a clean room in Sriharikota, before mounting on the launch rocket is shown in Fig.~\ref{fig:spacecraft}.  AstroSat was designed for an operational life of at least 5 years, and  is currently in the 7$^{{\rm th}}$ year of operations with its multiple instruments still operating and providing useful scientific data.

\section{\textit{AstroSat: Configuration and Auxiliary instruments}}

\begin{figure}[h]
\centering
\includegraphics[width=120mm]{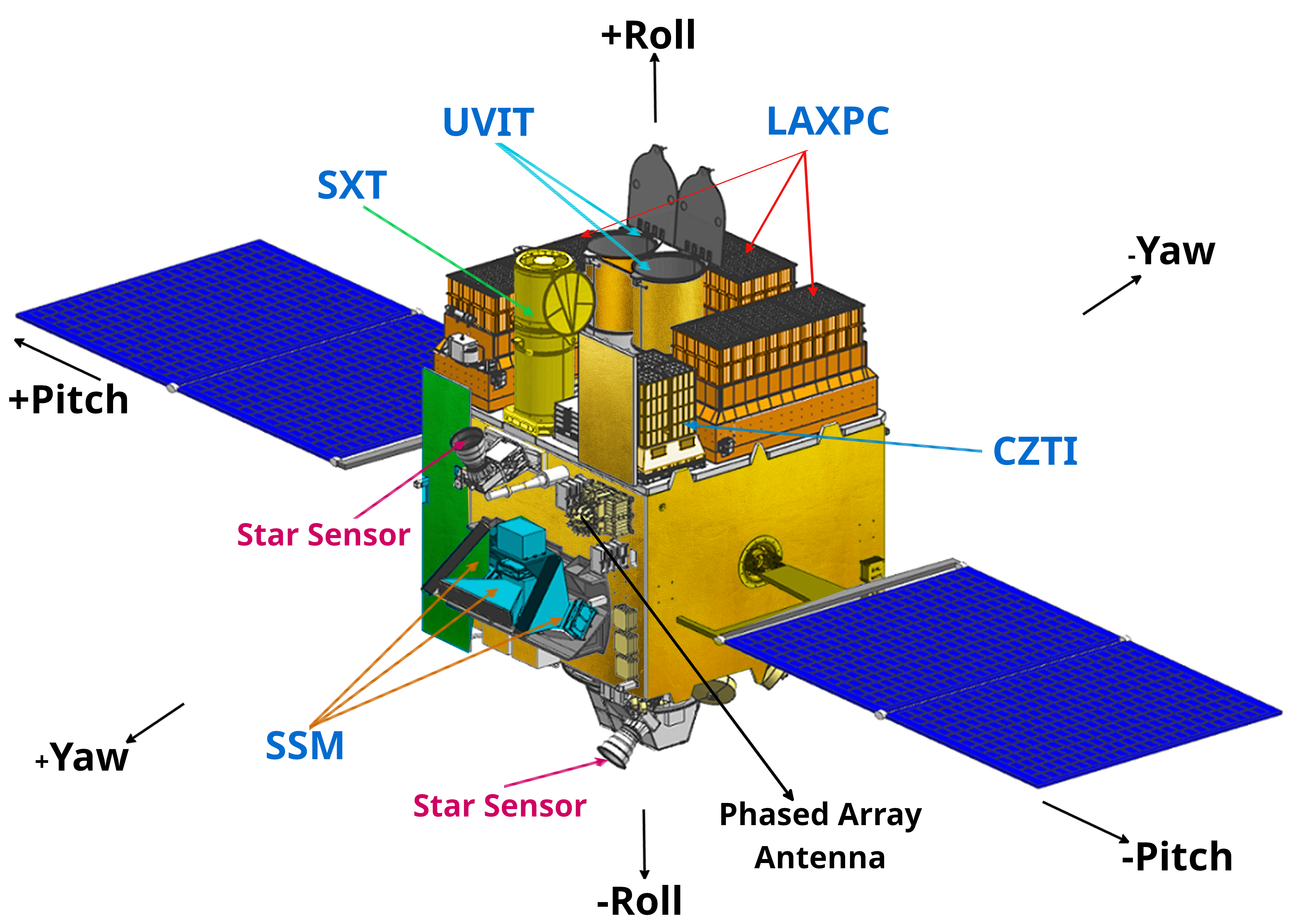}
\caption{AstroSat satellite schematic (Credits: V. Girish, ISRO HQ).}
\label{fig:schematic}       
\end{figure}

A schematic diagram of AstroSat showing its configuration is given in Fig.~\ref{fig:schematic}.  The four co-aligned scientific instruments have their view axis along the positive roll axis as the entire satellite can be rotated around this axis.  Two solar panels are mounted along an axis known as the pitch axis which is perpendicular to the roll axis.  The other axis perpendicular to both the roll and pitch axes is called the yaw axis. A brief description of its auxiliary systems is as follows:

\subsection{\textit{The Attitude and Orbit Control System (AOCS)}}

AstroSat is stabilised along all three axes and can be kept oriented towards any source in the sky,  while maintaining the pointing direction for the duration of an observation using the AOCS.  
It employs four reaction wheels (10 NMs) and three magnetic torquers of 60 Am$^2$ capacity.  A typical observation can last from a fraction of an hour to several days continuously.  A back up system of tiny jets (8 x 11 N hydrazine catalytic thrusters) is also available for orientation, but is only used in case of emergency so as to avoid the possibility of contamination of the optical elements of telescopes. Two star sensors, operating in a closed loop along with gyro wheels, control the attitude of the satellite and maintain it within $\sim$2 arcmin of the pointing direction.
Finer attitude determination is provided by the visible channel of one of the UVIT telescopes.  A pointing accuracy of $<$0.05 arcsec in each axis, an attitude drift rate of $\sim$1.1 arcsec s$^{-1}$ and a jitter of $<$0.3 arcsec is thus obtained.  It is required that the Sun should not to be in the field of view of any of the payloads for safety reasons.  Therefore,  the satellite pointing is always such that most of the time the Sun is around the negative yaw axis while the radiator plates for the SXT and CZTI detectors are on the opposite side.  Two antennae of a Satellite Positioning System (SPS) onboard AstroSat provide continuous position vector information.  
The targeting and movements between different targets is managed by a Bus Management Unit (BMU) such that the view axis of the instruments is always away from the Sun $(\geq 65$ degree), the limb of the Earth $(\geq$12 degree), and to keep the RAM angle (the angle between the roll axis and the velocity vector direction of the spacecraft) $\geq$12 degree.

\subsection{\textit{Timing Information}}

The timing signal for all the instruments is provided by the GPS (Global Positioning System) time interface to the LAXPC payload and the SPS. 
	
\subsection{\textit{Power Source}}	
	
The solar panels are made of triple junction solar cells. The panels are kept oriented in a direction normal to the Sun in order to generate maximum power of $\sim$2.2 kW.  Each solar panel has a size of 1.4 m x 1.8 m and is 20 mm thick. The panels can be rotated independently of the satellite around the pitch axis.  The panels were deployed after the satellite was injected into its orbit.  Two Li-ion batteries of 36 AHr capacity serve as a backup source of power when the orientation is such that the solar panels are not working at full efficiency. Power to the payloads is provided by a power management system.

\subsection{\textit{Thermal Control}}	
	
Space qualified thermal control elements such as optical solar reflectors, multi-layer insulators, active thermal control heaters, quartz wool blankets, titanium based thermal isolators, sink plates and ammonia heat pipes embedded in the satellite decks are used to maintain AstroSat within a comfortable temperature range of 0 to 40 degrees Centigrade.  There are 245 thermistors, 62 fine temperature sensors and 791 heater elements onboard AstroSat.   The scientific instruments are maintained within a temperature range as specified by the scientists and engineers, while the satellite goes through extreme hot (day) and cold (night) cycles as its passages around the Earth every 98 minutes.  The LAXPC is maintained between 10 degree C and 35 degree C, the UVIT telescope between 17 degree C and 23 degree C, and the optics module of the SXT  between 15 degree C and 25 degree C.  Two radiator plates -- one for cooling the CZTI detectors and the other for the thermo-electric cooler on which a CCD is mounted inside the SXT Focal Plane Camera Assembly (FPCA) are also used.  The radiator plates are connected to the payload via heat pipes using ammonia in the case of CZTI and ethane in the case of SXT.

\subsection{\textit{The Bus Management Unit}}	
	
The Bus Management Unit (BMU) integrates the main functions of AstroSat, including the AOCS, telemetry, command and sensor processing, etc. 
	
\subsection{\textit{Data Storage and Handling}}
	
A solid-state recorder with 200 Gb storage capacity is used for on-board storage of data. Data from all the payloads are transmitted to the Mission Operations Complex, ISRO Telemetry Tracking and Command Network (ISTRAC) by two X-band carriers using phased array antennae.  The same station operates the satellite. All data are transmitted in 11 or 12 orbits (out of the $\sim$14 orbits per day) over a day generating $\sim$700 GB of scientific data every month. Data are processed, distributed and archived by the Indian Space Science Data Centre (ISSDC) near Bengaluru for all the observers. 

\subsection{\textit{Communications Systems}}	
	  	
The communication system for AstroSat mission consists of the Telemetry and TeleCommand (TTC) transponder systems in the S-band and data transmission systems which use two X-band carriers with steerable Phased Array Antenna systems. The data are transmitted at a rate of 105 Mbps, once in all the visible orbits.
\\
\\
For more details of the auxiliary systems please see \cite{Navalgund2017} and \cite{Pandiyan2017}.

\section{\textit{Choice of Orbit}}

The orbital inclination of 6 degrees was selected to avoid the satellite going through the inner parts of the South Atlantic Anomaly region with a concentration of charged particles that lead to very high backgrounds in the instruments. The altitude of 650 km also minimises the effect of atomic oxygen on the optical elements of the UVIT and SXT, and other components of the satellite.  The satellite is eclipsed by the Earth for $\sim$37 minutes each revolution and its visibility from the ground station at Bengaluru is between 5 to 12 minutes.

\section{\textit{Scientific Payload}}

The scientific payload onboard AstroSat consists of the following instruments:

\subsection{\textit{Ultraviolet Imaging Telescopes (UVIT)}}

\begin{figure}[h]
\centering
\includegraphics[width=120mm]{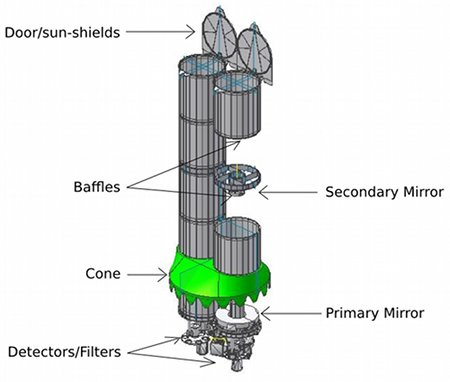}
\caption{UVIT Telescopes (Credits: UVIT Team at IIA and ISRO).}
\label{fig:uvit}       
\end{figure}

A schematic diagram of the Ultra Violet Imaging telescopes (UVIT) is shown in Fig.~\ref{fig:uvit}. The instrument consists of twin telescopes with a total mass of 201 kg.  Each telescope is 3.1 m long and has diameter of 0.877 m.  One of the telescopes is for the Far UV (FUV) wavelength range (130 –- 180 nm), and the other for both Near-UV (NUV) (200 -– 300 nm) and visible (VIS) (320 -– 550 nm) wavebands.  In the NUV/VIS telescope, a beam splitter is used to reflect NUV and transmit VIS.   
Each telescope has a hyperbolic f/4.5 primary mirror of 375 mm optical diameter, and a hyperbolic secondary mirror of 140 mm optical diameter. The overall focal length of the telescopes is 4.75 m with a focal ratio of f/12 in Ritchey–Chretien optical configuration.  The reflective coatings are Al with MgF$_2$ layer with a surface reflectivity of $\sim$70$\%$ in FUV, and $>80\%$ in the NUV and VIS. The roughness of the coated surfaces is $<$1.5 nm $rms$, to keep the scattering $<1\%$ per surface.   
The primary use of the VIS channel is as a very fine star tracker to obtain the aspect of the UVIT telescopes with a sufficiently high accuracy.  The field of view (FoV) of each telescope is 28 arcmin.  A host of filters and low resolution gratings are mounted in a filter wheel in front of the detectors in their respective focal planes allowing a choice of narrower wavelength bands.  The wheels for the FUV and NUV channels are also equipped with gratings for low resolution slitless spectroscopy.  A cylindrical baffle mounted in front of the telescope helps to avoid stray light from bright off-field sources with an attenuation of a billion (10$^9$)  for sources that are 45 degree away from the axis of the telescope.  A door on top of each telescope, which protects the telescopes from contamination on ground, is opened after launch and then kept open. 
The doors also act as Sun shades provided the Sun is more than 45 degree from the viewing axis. The bright-earth is kept away from the view axis by $>$12 degree and the Sun is kept behind the Sun-shade at all times even if UVIT is not observing, to avoid contamination of the optics due to ultraviolet assisted reactions.  In addition, the nominal observation period is restricted to the night time to avoid contamination from scattered geo-coronal lines, scattered by the other instruments on the spacecraft into the baffles.  All the optical components of the UVIT, e.g., telescope tubes, telescope rings, detector mounting brackets, spider rings etc are made of Invar36, while the other parts are made of an aluminium alloy.   The two telescopes are held together onto the satellite deck using a cone-like interface made of titanium material.  Active thermal control is provided to reduce effects of temperature variations in the orbit on the optics, e.g., defocusing etc.

\begin{figure}[h]
\centering
\includegraphics[width=0.54\textwidth,clip]{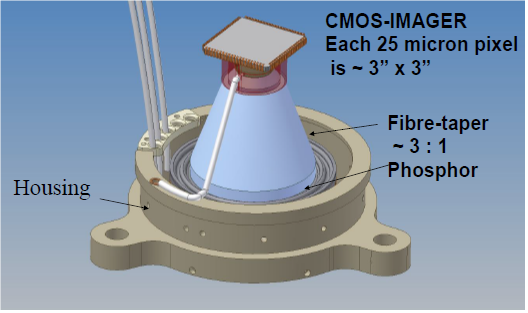}
\includegraphics[width=0.45\textwidth,clip]{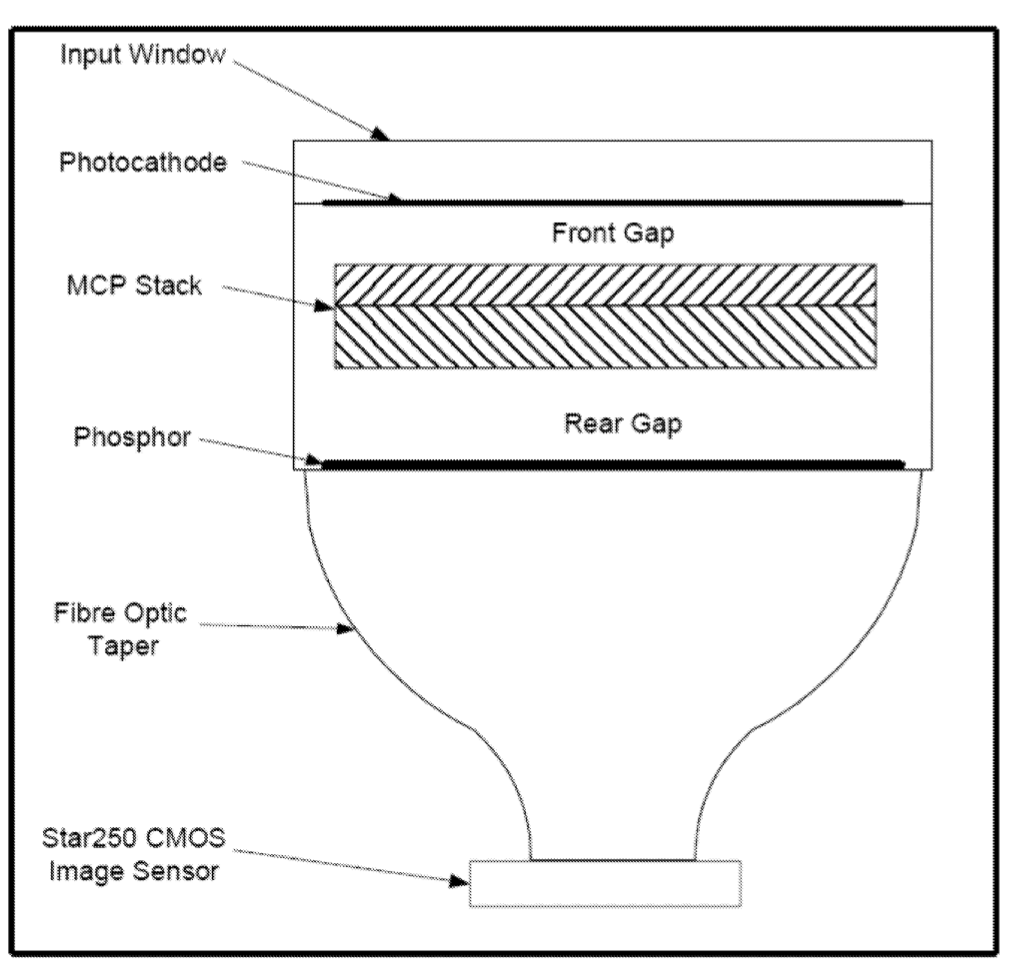}
\caption{UVIT Detectors (Credits: UVIT Team at IIA).}
\label{fig:uvitdetector}       
\end{figure}

The camera in the focal plane of each of the UVIT telescopes has intensified imagers to detect incoming photons.  These cameras have an aperture of 39 mm, and consist of intensified CMOS (Complementary metal oxide semiconductor).   A schematic diagram of one of the detectors used in the UVIT is shown in Fig.~\ref{fig:uvitdetector}.   The three detector systems in each channel are identical except for the window and the photo cathode used. The  window in the FUV detector is made of MgF$_2$ and the photo cathode is made of CsI. The windows in the NUV and VIS detectors are made of silica, and the photocathodes are made of CsTe and S20 respectively.   The incident photon produces electrons in the photocathode which are accelerated across a gap of $\sim$0.1 mm onto a stack of two crossed microchannel plates (MCP) which have a gain of $\sim$10$^7$ , resulting in a shower of electrons illuminating a phosphor with a fast decay time. The light from the phosphor is channeled through a tapered structure made of fiber-optics, on to a Fillfactory/Cypress Star250 CMOS detector with 512 x 512 pixels with each pixel of 22 microns size on a side. The CMOS detector serves as the readout system of the detector.   Each of these pixels is mapped to 8 x 8 subpixels in the final image to get a plate-scale of $\sim$0.416 arcsec per subpixel.  The taper structure is bonded to the surface the CMOS detector.  There is a loss of about a factor of 10 from the light produced in the phosphor in the process.  

The splash of light produced by an incident photon and registered on the CMOS that covers several pixels is called an event.  An FPGA (field-programmable gate array) recognises each event and performs centroiding in real time.  The centroid is determined by finding a pixel with a value that is larger than any of its 8 surrounding pixels. The total signal for an event is calculated using a range of 5 x 5 pixels surrounding this central one.  The gap between the photocathodes and the micro channel plates is $\sim$0.1 mm to minimise lateral movement of the photo-electrons and get a spatial resolution of $<$0.025 mm FWHM, in the FUV and NUV channels.  The detectors can work in a photon counting (PC) mode or in an integration mode.  The time resolution in the PC mode is 1.7 msec.  Integration mode is used when the source is bright and multiple photons are expected per readout frame.  Photon counting mode is used for weak sources with total expected count rate of only $\sim$1 or less per frame.  In the PC mode, the signals above a threshold are also called photon events.  In the PC mode if a point source gives multiple photons in a frame these are also counted as one. This loss of photons is corrected by invoking Poisson statistics as long as the average rates are up to $\sim$2 photons per frame.  At higher rates the average signal for individual photons is reduced due to impedance of the micro channel plates.  The loss of photons due to this effect is expected to be $<5\%$ for rates up to 150 events s$^{-1}$.  For images containing the entire field of view the maximum frame-rate can be $\sim$29 s$^{-1}$.  Images of partial windows of 100 x 100 pixels corresponding to a field of $\sim$5.5 arcmin x 5.5 arcmin on the CMOS can have event rates of up to 600 s$^{-1}$, thus allowing much brighter objects to be seen.  The satellite pointing axis drifts up to ($\pm$)1 arcmin with a drift rate of $\sim$3 arcsec s$^{-1}$. The recording rate of 29 Hz or higher keeps the blur in individual images to less than 1 arcsec.  The drift is measured via images taken by the VIS channel or the NUV channel every second, as most fields are very faint in the FUV. The final images are made at the receiving station on ground using a shift and add algorithm.

\subsubsection{\textit{UVIT Filters}}

The characteristics of a large number of filters used in front of the detectors are summarised in Table~\ref{tab:uvitfilters}, where $\lambda$ is the wavelength, $\Delta \lambda$ is the bandwidth, UC is Unit Conversion for the filter; and ZP is the Zero point for the filter at the centre of the field.

\begin{table}   
\begin{center}
\begin{tabular}{| l | l l l l l l l |} \hline
	
Channel & Slot & Filter & Name & Mean $\lambda$ & $\Delta \lambda$ & ZP & UC \\

        &      &        &      &  (nm)          & (nm)             & AB mag & $10^{-15}$ erg s$^{-1}$ cm$^{-2}$ ${\rm \AA}^{-1}$ \\

\hline

FUV & 0 & Block & Block &&&&\\
& 1 & F148W & CaF2-1 & 148.1 & 50.0 & 18.097 $\pm$ 0.01 & 3.09 $\pm$ 0.029 \\
& 2	& F154W	& BaF2	& 154.1	& 38.0	& 17.771 $\pm$ 0.01	& 3.55 $\pm$ 0.04 \\
& 3	& F169M	& Sapphire & 160.7 & 29.0 &	17.41 $\pm$ 0.01 & 4.392 $\pm$ 0.037 \\
& 4	& Grating-1 &&&&&\\				
& 5	& F172M & Silica & 171.6 & 12.5 &16.274 $\pm$ 0.02 &10.74 $\pm$ 0.16 \\
& 6	& Grating-2 &&&&&\\				
& 7	& F148Wa &	CaF2-2 & 148.5 & 50.0 & 18.008 $\pm$ 0.01 & 3.28 $\pm$ 0.025 \\
NUV	& 0 & Block	& Block &&&&\\				
&	1 &	N242W &	Silica-1 & 241.8 & 78.5 & 19.763 $\pm$ 0.002 & 0.222 $\pm$ 0.00065 \\
&	2 & N219M	& NUVB15 & 219.6 & 27.0 & 16.654 $\pm$ 0.02 & 5.25 $\pm$ 0.082 \\
&	3 & 0N245M &NUVB13 & 244.7 & 28.0 & 18.452 $\pm$0.07 & 0.725 $\pm$ 0.0036 \\
&	4 &	Grating	&&&&&			\\
&	5 & N263M & NUVB4 & 263.2 & 27.5 & 18.146 $\pm$ 0.01 & 0.844 $\pm$ 0.0096\\
&	6 & N279N & NUVN2 & 279.2 & 9.0 & 16.416 $\pm$ 0.01 & 3.50 $\pm$ 0.035 \\
&	7 &N242Wa & Silica-2 &&&&				\\
VIS &	0 &	Block &	Block &&&&\\				
&	1 & V461W & VIS3 & 461.4 & 130 & NA & NA \\
&	2 & V391M & VIS2 & 390.9 & 40 & NA & NA \\ 
&	3 & V347M & VIS1 & 346.6 & 40 & NA & NA \\
&	4 & V435ND & ND1 & 435.4 & 160 & NA & NA \\
&	5 & V420W & BK7 & 420.0	& 220 & NA & NA \\

\hline
\end{tabular}
\caption{{UVIT Filter characteristics. See \cite{Tandon2017a}, \cite{Tandon2017b} and \cite{Tandon2020} }}
\label{tab:uvitfilters}
\end{center}	
\end{table}

For an observed count rate or counts per sec (CPS), the source Flux (F) and its AB magnitude for a filter can be estimated using the following relations: 

\begin{equation}			
	 {\rm F}_{\rm filter} = {\rm CPS \ x \ UC \ (erg \ s}^{-1} \ {\rm cm}^{-2} \ {\rm \AA}^{-1})  
\end{equation}

\begin{equation}			
	 {\rm m_{AB} = -2.5 \ log_{10}(CPS) \ + \ ZP}	
\end{equation}
 
\begin{figure}[h]
\centering
\includegraphics[width=0.48\textwidth,clip]{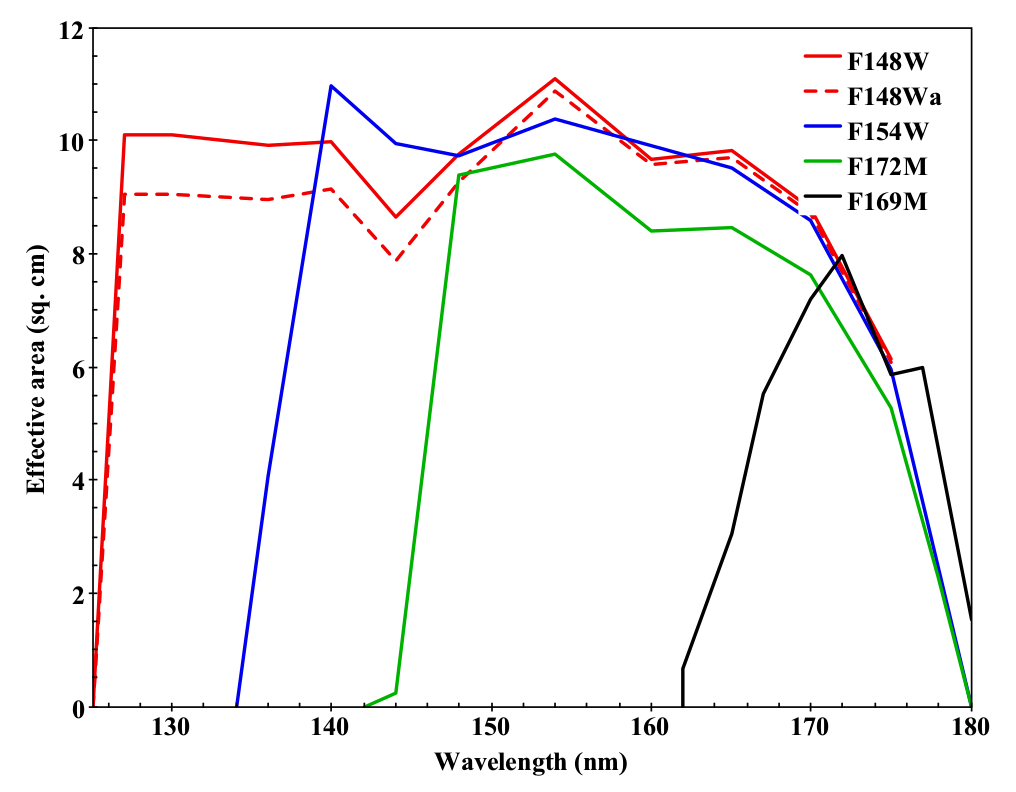}
\includegraphics[width=0.48\textwidth,clip]{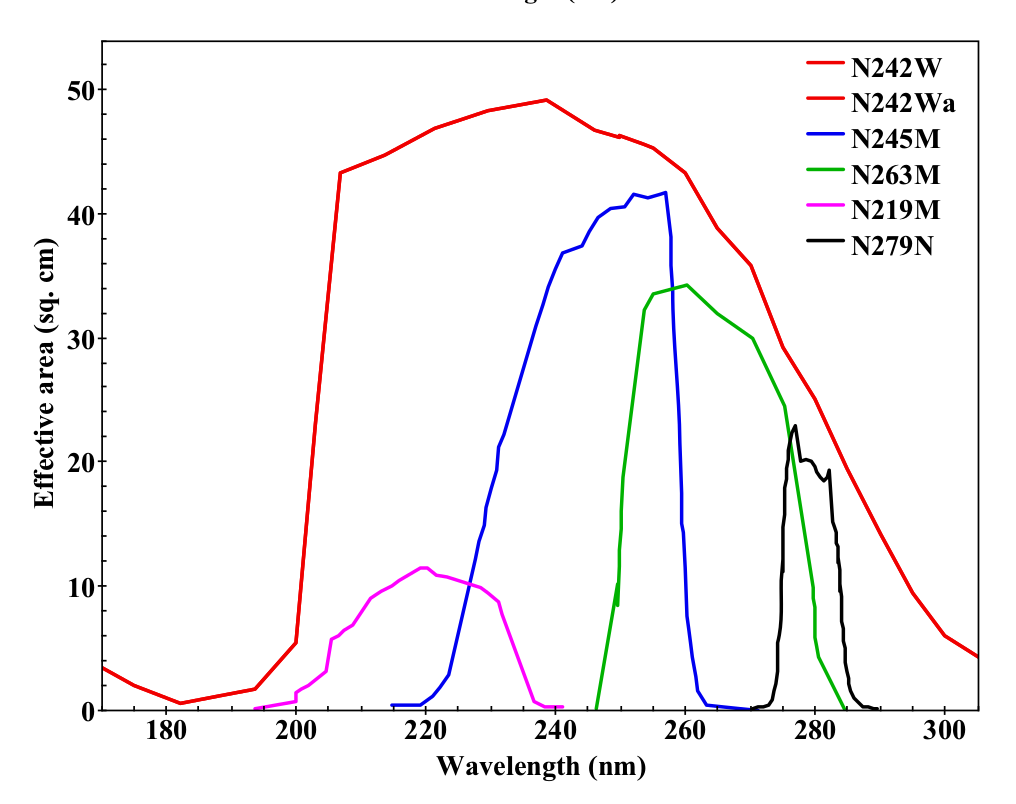}
\includegraphics[width=0.48\textwidth,clip]{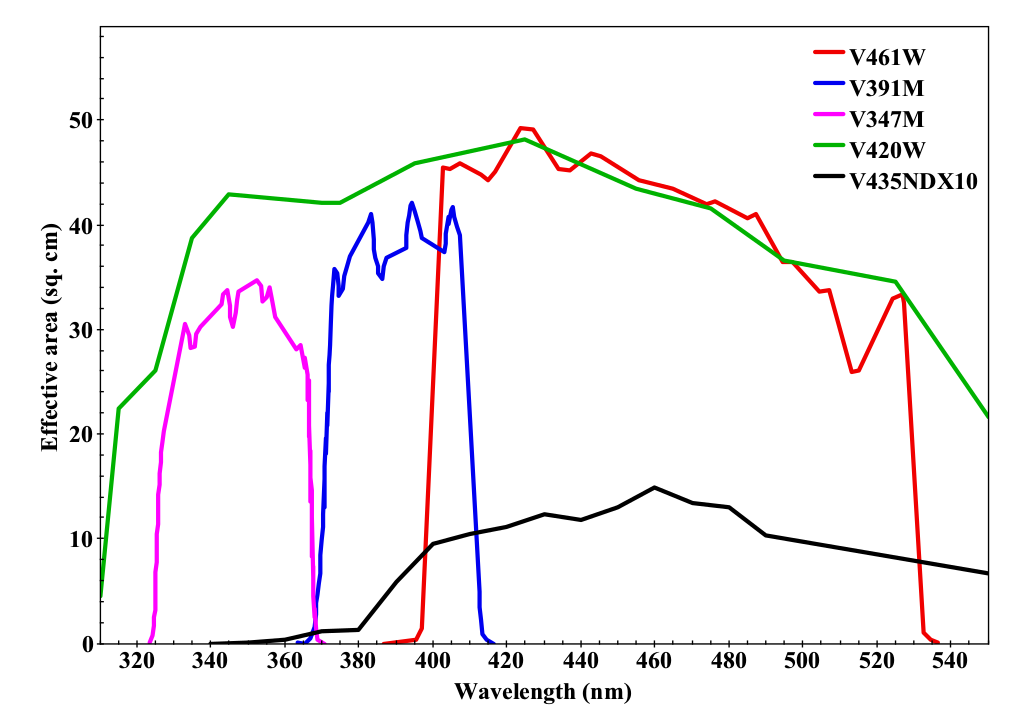}
\caption{Effective Areas of UVIT with different filters (Credits: UVIT Team at IIA).}
\label{fig:uviteffarea}       
\end{figure}

The effective areas for the different channels with different filters are shown in Figure~\ref{fig:uviteffarea} (FUV -- top left), (NUV -- top right), and VIS (bottom) as a function of wavelength.  The spatial resolution as measured in orbit is found to be $\sim$1.2 arcsec (FWHM of the point spread function – PSF) for NUV and $\sim$1.5 arcsec for FUV  (significantly better than the goal of 1.8 arcsec before launch) in the PC mode.   The PSF is a factor of $\sim$3 better than that of GALEX which had a much larger foV ($\sim$1.20 degree diameter), however.  The effect of temperature change on the focus is $<0.032$ mm/degree C.  The contribution of defocus to the PSF is, therefore,  $<$0.15 arcsec (rms) for a variability of $<$3 degree C in the temperature in an orbit. The relative alignment between the centres of the FUV and NUV fields have been measured to be within 70 arcsec, while in orbit. 

The use of intensified imagers can lead to permanent damage in UVIT detectors if exposed to excessive flux of photons. Therefore, observations are only allowed for those fields which have objects below defined levels of brightness in the 40 arcmin diameter field around the centre taking into account possible pointing error which can be up to 3 arcmin and some drift during the exposure.  The following upper limits on count rates are used for a point source (or any 100 arcsec$^2$ part of an extended source) in the field: a) for VIS-CPU: 4800 s$^{-1}$ (in integration mode with an intensification $\sim$1/20 of that used for photon-counting), b) for NUV-CPU: 1500 s$^{-1}$ (in PC Mode) and c) for FUV-CPU: 1500 s$^{-1}$ (PC Mode). A total rate of 10$^5$ s$^{-1}$ over the full face of the detector is considered safe in the PC mode.

\begin{figure}[h]
\centering
\includegraphics[width=120mm]{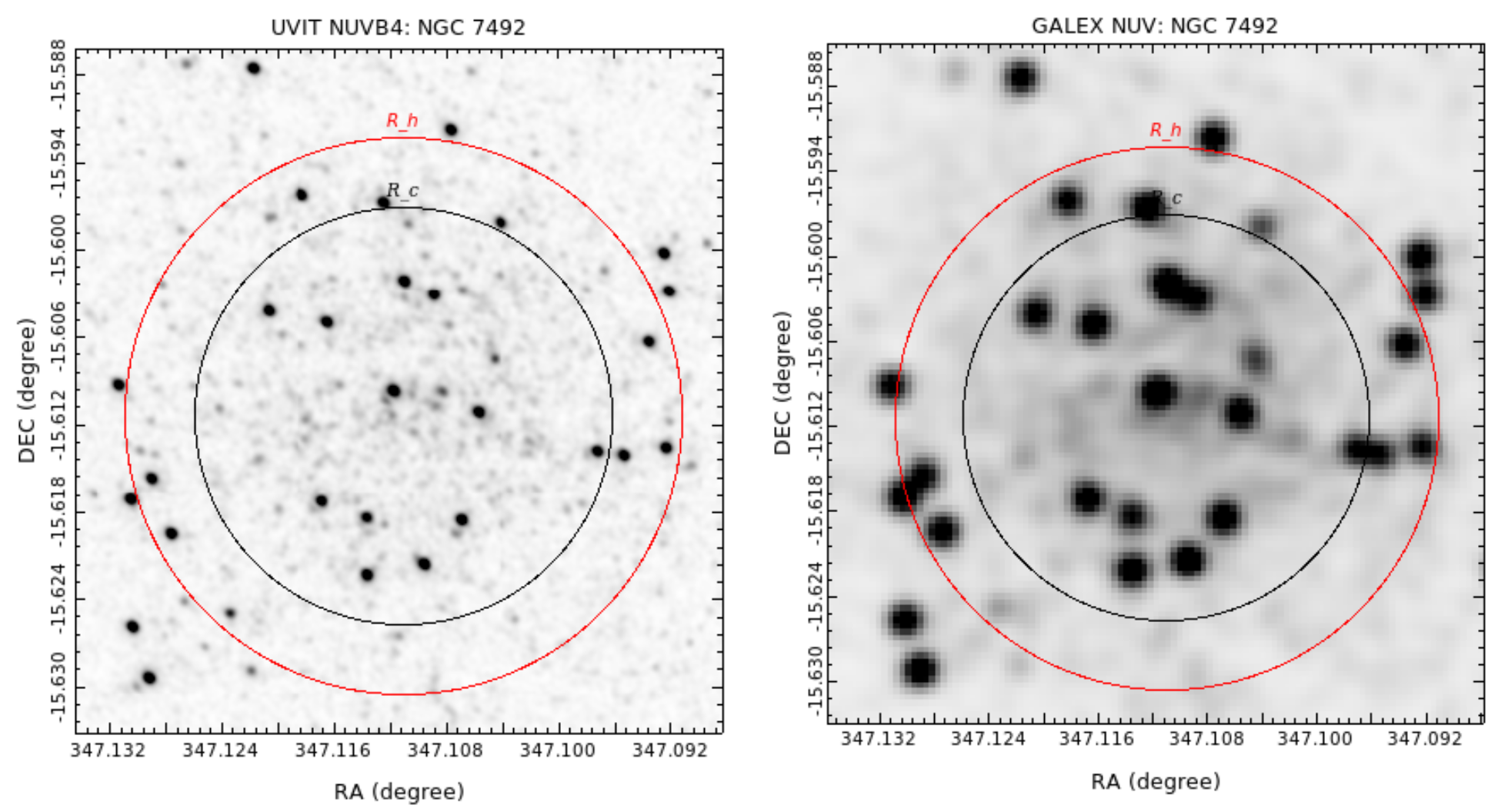}
\caption{UVIT N263M image (left) and GALEX NUV (eff. wavelength = 230.4 nm, right) images of a Globular cluster NGC 7492. The red circle denotes the half light radius of 1.15 arcmin and the black circle denotes the core radius of 0.86 arcmin) of the cluster. Credits:~\cite{Kumar2012}.}
\label{fig:uvitimage1}       
\end{figure}

\begin{figure}[h]
\centering
\includegraphics[width=0.49\textwidth,clip]{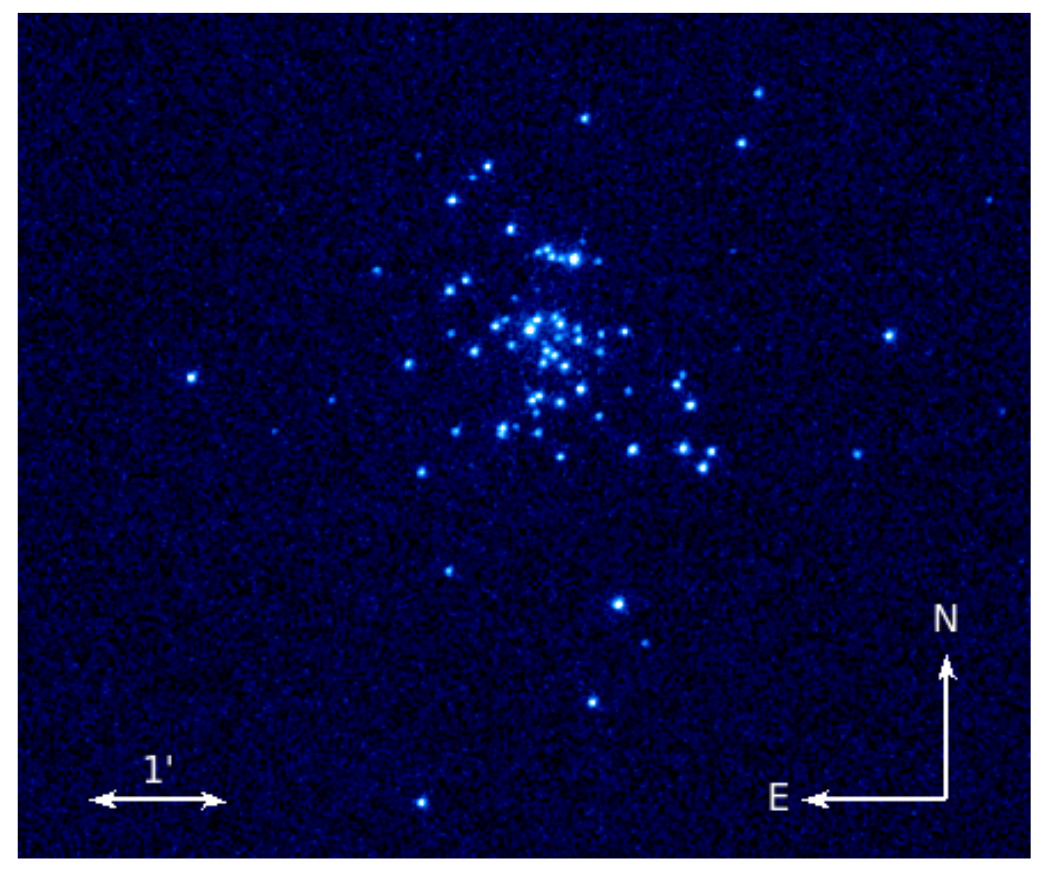}
\includegraphics[width=0.49\textwidth,clip]{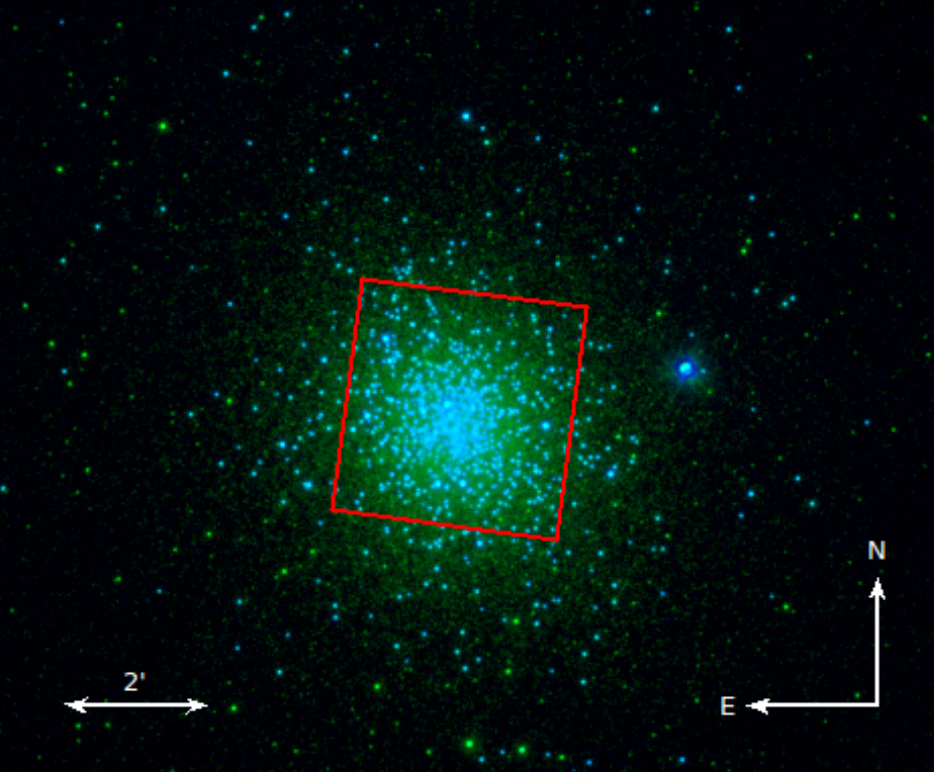}
\caption{(Left) Globular cluster NGC 2298 imaged in UVIT F148W filter (credits: \cite{Rani2021}). (Right) Colour composite image of Globular cluster NGC 2808 in F154W filter shown in blue and N242W filter shown in green.  The Hubble Space Telescope Wide Field Camera which covers the inner 2.7 x 2.7 arcmin$^2$ region of the cluster is marked in red. (Credits:~\cite{Prabhu2021}).}
\label{fig:uvitimage2}       
\end{figure}

\begin{figure}[h]
\centering
\includegraphics[width=0.58\textwidth,clip]{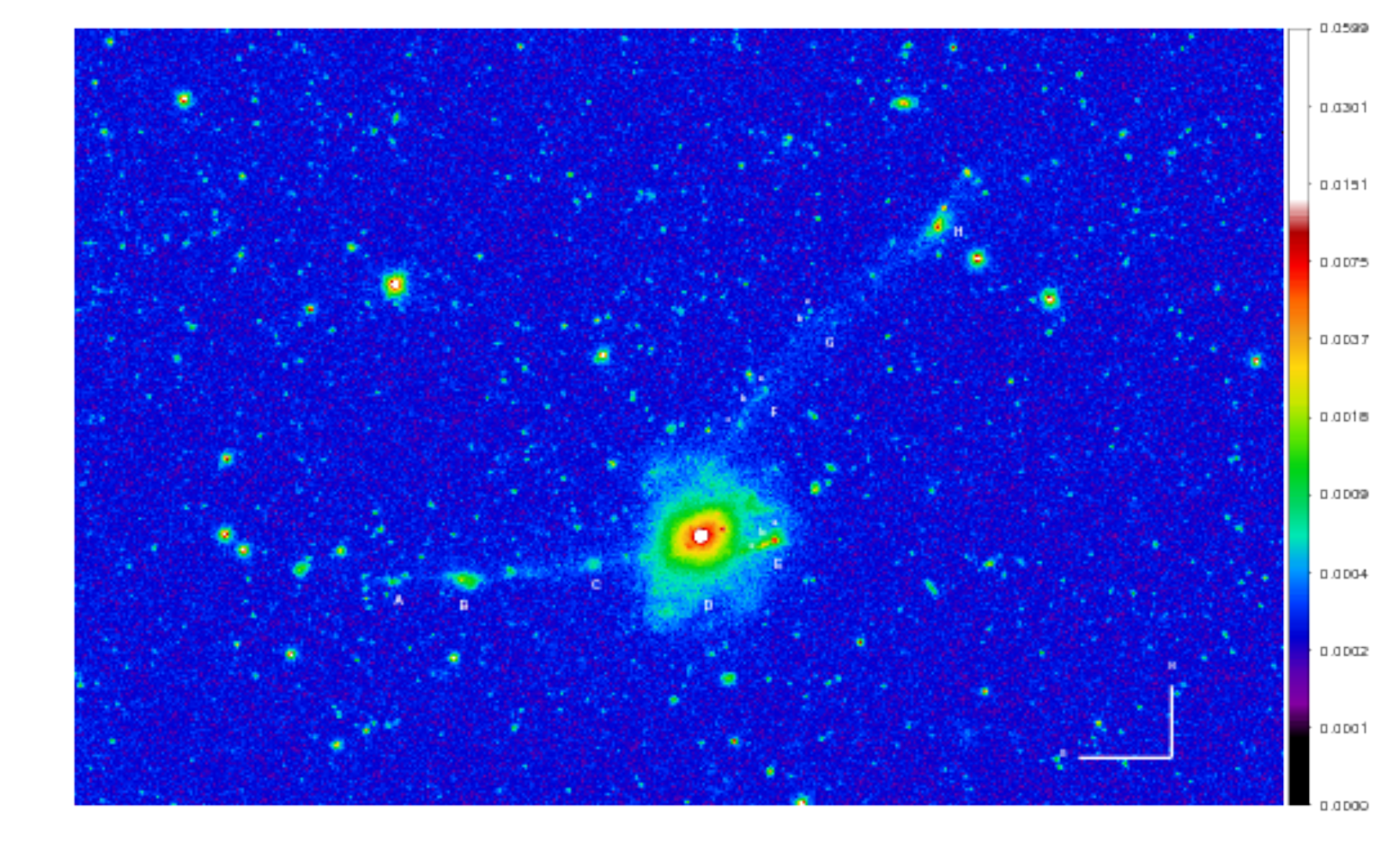}
\includegraphics[width=0.41\textwidth,clip]{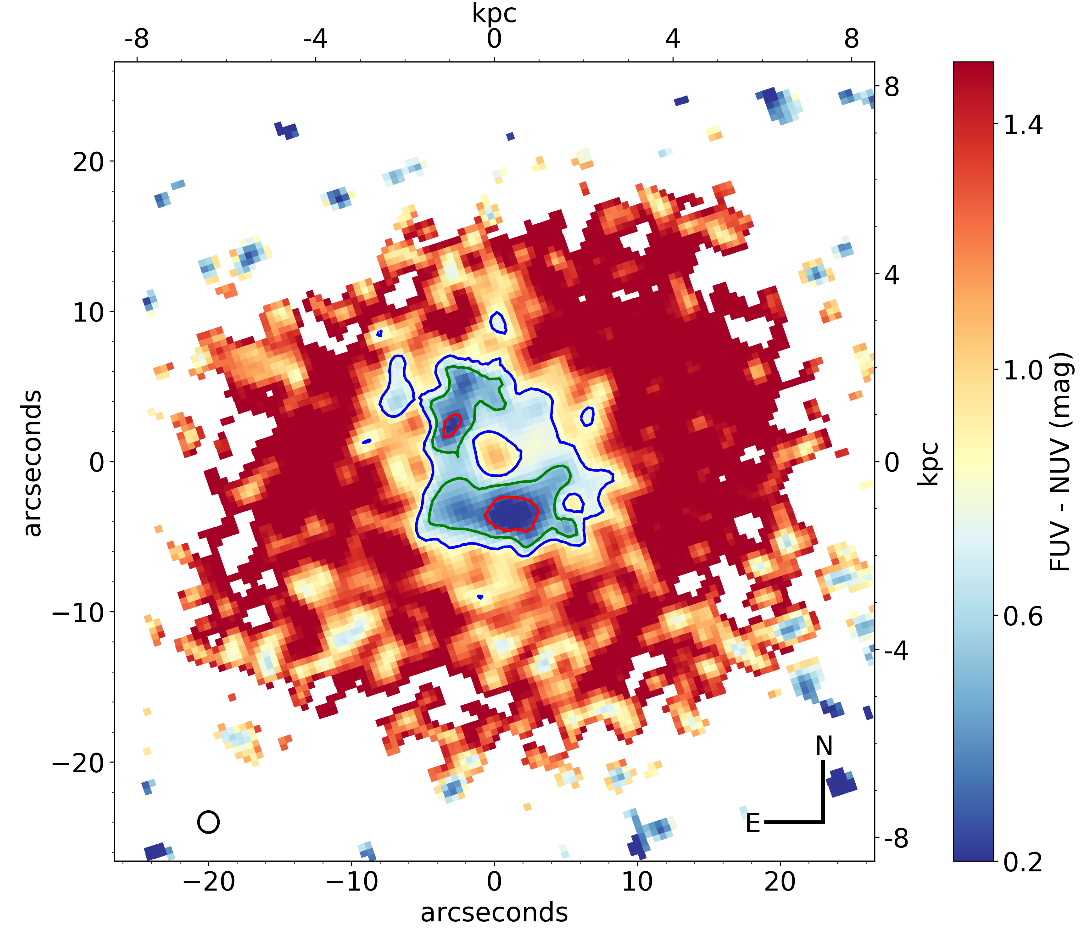}
\caption{(Left)The NUV image of post-merger galaxy NGC 7252 showing the low surface brightness tidal tails extending up to 100 kpc. (Right) FUV-NUV color map of the central parts of NGC 7252 (size $\sim$16 kpc on a side. Contours represent star forming regions of age of 150 (red contour), 250 (green contour), 300 (blue contour) Myr to isolate regions of constant age. The blue colour ring in the centre hosts young (150 Myr) stars compared to the rest of the galaxy. (Credits: \cite{George2018a}, \cite{George2018b}).}
\label{fig:uvitimage3}       
\end{figure}

The FUV channel can detect a 20 magnitude (AB magnitude system) star at 5$\sigma$ level in $\sim$160 s exposure. The NUV channel worked only for the first 3 years after launch. 
Some of the images obtained with the UVIT are shown in Figs.~\ref{fig:uvitimage1}, \ref{fig:uvitimage2} and \ref{fig:uvitimage3}, to illustrate the quality of imaging achieved. A comparison with the images produced by GALEX is also shown in Fig.~\ref{fig:uvitimage1}.

\subsubsection{\textit{UVIT Gratings}}

\begin{figure}[h]
\centering
\includegraphics[width=120mm]{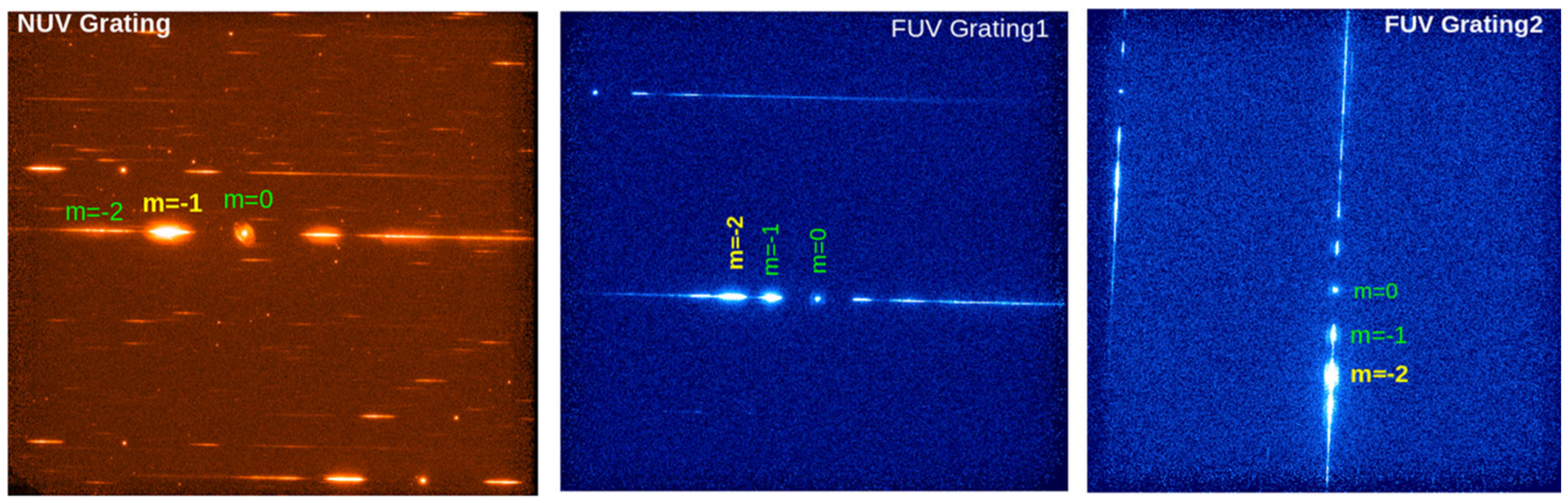}
\caption{NUV and FUV grating images of NGC40. The negative grating orders are marked. The image sizes are 9.8 arcmin on a side (Credits:~\cite{Dewangan2021}).}
\label{fig:gratingimages}       
\end{figure}

UVIT carries three gratings: one mounted on the NUV channel filter wheel while the other two are mounted on the FUV channel filter wheel such that their dispersion axes are nearly orthogonal to each other to obviate possible contamination from nearby sources in the grating images.  A nearby source along the dispersion arm of the main target will cause contamination in one FUV grating image while in the other FUV grating image the dispersion arms will be well separated.  The gratings are ruled with 400 lines mm$^{\rm -1}$ on CaF 2 substrates of 4.52 mm thickness.  The dispersion in the detector plane by a grating is 1.2 nm arcsec$^{-1}$ and 0.6 nm arcsec$^{-1}$ in the first and second order, respectively, at 135 nm.    The detection efficiency of the FUV gratings is maximum for the m = -2 order while the NUV gratings have maximum efficiency for the m = -1 order.  The spectral resolution for the FUV gratings is 1.46 nm and 3.3 nm (FWHM) for the NUV gratings. The PSF is somewhat broader (FWHM $\sim$2 arcsec) for the gratings. The gratings have been calibrated using observations of NGC 40, a planetary nebula rich in emission lines well-suited for wavelength calibration (see Fig.~\ref{fig:gratingimages}), and HZ 4 –- a white dwarf HZ 4 with almost featureless spectrum appropriate for flux calibration and for cross-calibration with the wide-band filters.

The observed emission line positions in the grating spectra of NGC 40 were fitted with Gaussian profiles plus a low order polynomial for the continuum thus identifying the strong emission lines.  The line wavelengths and the corresponding pixel numbers were fitted with a linear relation. The coefficients of the linear dispersion relation are given in Table 2 taken from \cite{Dewangan2021}.  

\begin{table}   
\begin{center}
\begin{tabular}{| l l l l l l |} \hline
	
Grating & Tilt Angle & Order & Order Range & Intercept & Slope  \\
        & ($\theta$) degrees &  & in relative pixel & & \\
        &                    &  & coordinate (X) & & \\

\hline
FUV-G1 & 358.70 & -2 & -629 to -413 & 43.4 & -2.791 \\
       & 358.70 & -1 & -323 to -213 & -18.0 & -5.833 \\
FUV-G2 & 267.53 & -2 & -624 to -426 & 31.2 & -2.812 \\
       & 267.53 & -1 & -313 to -228 & 45.0 & -5.625 \\
NUV-G  & 358.90 & -1 & -545 to -336 & 45.1 & -5.523 \\      

\hline
\end{tabular}
\caption{{Linear dispersion Relations for the UVIT Gratings (Ref.~\cite{Dewangan2021})}}
\label{tab:uvitgratings}
\end{center}	
\end{table}
 
\subsubsection{\textit{UVIT Analysis Software}}

A graphical user interface (GUI) software known as CCDLAB has been developed for the Windows operating system for basic data reduction and analysis of images obtained with the UVIT.  CCDLAB is written in Visual Studio Professional 2008, using C$++$. The GUI can be used to extract and process the Level1 (L1) data provided by ISRO to the UVIT Payload Operations Centre (POC) at IIA, Bengaluru for verification and then posted at the “Science Data Archive for AstroSat Mission” accessible from 
\url{https://astrobrowse.issdc.gov.in/astro_archive/archive/Home.jsp}. 
CCDLAB can perform various detector-specific and observational corrections to L1 data and to reduce it to scientific level data (Level 2 - L2) for further analyses.  For more details about CCDLAB please see \cite{Postma2017}. 
The official UVIT pipeline to convert L1 data to L2 data products is based on linux systems and is available for download with documentation and calibration database at \url{https://uvit.iiap.res.in/Downloads} (see \cite{Ghosh2021} for more details). The L2 data downloadable from the astrobrowse is produced by this pipeline. A python based software to produce light curves from the L2 data provided by the official pipeline is available at \url{https://github.com/prajwel/} (\cite{Joseph2021}).

More detailed information about the UVIT is given in \cite{Hutchings2007}, \cite{Postma2011}, \cite{Kumar2012a}, \cite{Tandon2017a}, \cite{Tandon2017b}, \cite{Tandon2020}), and \cite{Dewangan2021}.

\section{\textit{Large Area X-ray Proportional Counters (LAXPC)}}

\begin{figure}[h]
\centering
\includegraphics[width=120mm]{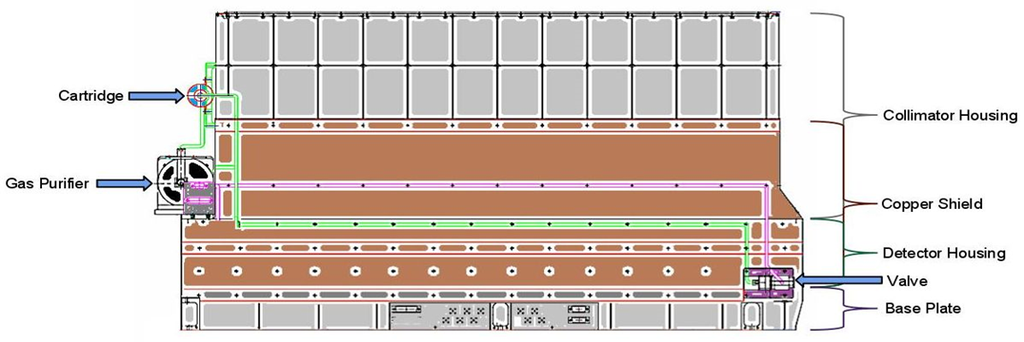}
\caption{A schematic diagram of one of the LAXPC detectors (Credits: LAXPC team at TIFR).}
\label{fig:LAXPC}       
\end{figure}

\begin{figure}[h]
\centering
\includegraphics[width=120mm]{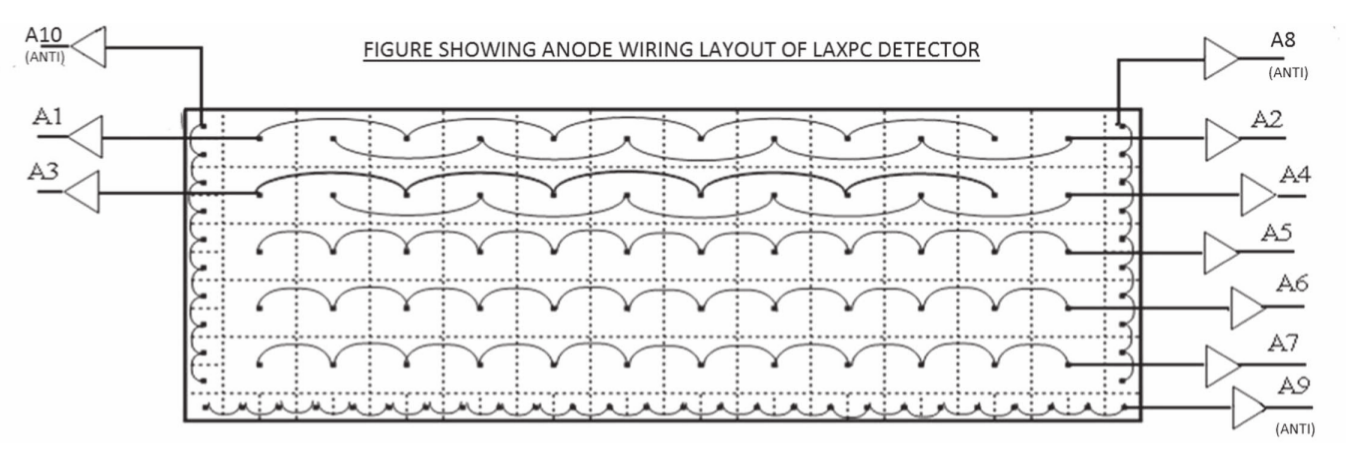}
\caption{A schematic of the configuration of anodes in a LAXPC unit (Credits: LAXPC team at TIFR).}
\label{fig:LAXPCanodes}       
\end{figure}

\begin{figure}[h]
\centering
\includegraphics[width=120mm]{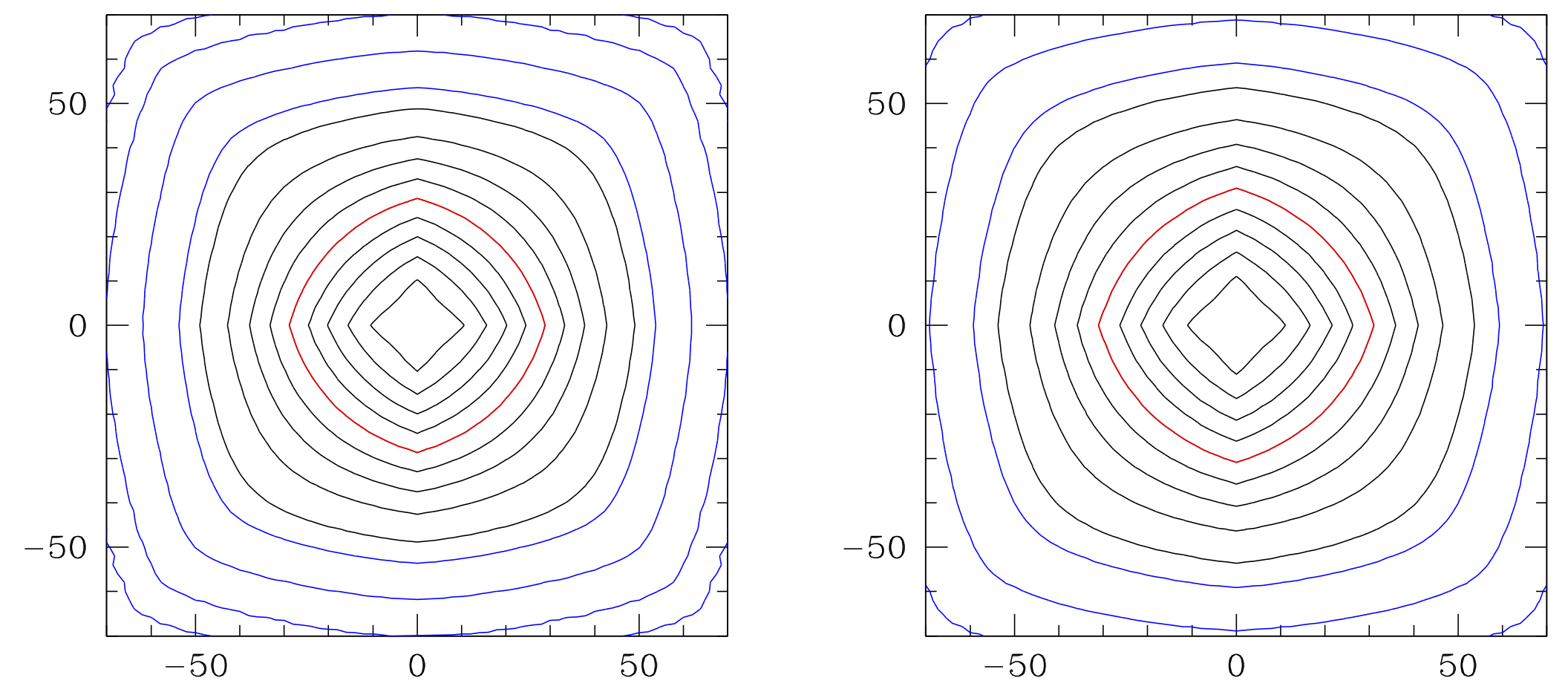}
\caption{Constant efficiency contours as a function of angle (relative to the detector axis) from the detector axis for photons of 15 keV (left panel) and 50 keV (right panel). The axes are in arcmin. The red contour encloses the region where count rate is more than half of the peak value and gives the FWHM of the field of view. The black contours are at intervals of 10$\%$, while the blue contours are at levels of 5$\%$, 1$\%$, 0.1$\%$, and 0.01$\%$. (Credits: \cite{Antia2017})}
\label{fig:LAXPCeffic}       
\end{figure}

This instrument consists of three (LX10, LX20, LX30) co-aligned, identical multi-layered and multi-cell proportional counters (PC) which detect X-rays in the 3 -– 100 keV energy range.  Each PC weighs 130 kg, has a length of 1.193 m, a width of 0.568 m and a height of 0.690 m.  A schematic diagram of one LAXPC unit including the collimators and shields is shown in Fig.~\ref{fig:LAXPC}.  Each unit has five layers with 12 anode cells with each cell being 100 cm long and having a cross-section of 3 x 3 cm, giving each PC a depth of 15 cm.  The top two layers are divided into two parts and alternate cells (odd and even cells) are connected for common high voltage and output signals. Thus, there are seven anodes in all - two each in the top and second layer, one each from the remaining three layers. Three Veto layers surround the left, right and bottom sides and serve to reduce background in the PC.  Each veto layer consists  of anode cells of 1.5 cm x 1.5 cm x 100 cm.  The outputs from the seven main anodes and three veto anodes are fed to 10 charge-sensitive pre-amplifiers (CSPAs). The anode configuration is shown in Fig.~\ref{fig:LAXPCanodes}. A common high voltage (2000 -- 2500 Volts) is supplied to all main anodes (A1 –- A7) at the same point ensuring the same voltage for the anodes. This voltage can be controlled from ground. The veto anodes have a lower voltage.  Two of the PCs, viz., LX10 and LX20 were filled with a mixture of xenon (90$\%$) and methane (10$\%$) gas at a pressure of $\sim$2 atmospheres, while the LX30 unit was filled with a mixture of xenon (84.4$\%$), methane (9.4$\%$) and argon (6.2$\%$) at  $\sim$2 atmosphere pressure. An onboard gas purification system is included in each PC, which is commanded from time to time to supply recycled gas after purification to maintain the properties like energy gain and resolution of the detectors close to the optimal value.  A 50$\mu$m thick aluminised Mylar sheet on top of each PC seals the gas, blocks X-rays with energy $<$3 keV, and thus dictates the low-energy threshold of the detectors.  A collimator made of slats sits on top of the Mylar window, and has $\sim$7000 square cells of 7 mm x 7 mm size.  The collimator serves as a window support structure and defines the field of view of 1 degree x 1 degree for X-rays incident on top of the collimator.  GEANT4 simulations were used to determine the FoV using 10$^6$ photons of fixed energy distributed uniformly over the top of the detector and incident at a fixed angle to the detector axis. The number of events registered in the detector was noted for each angle. Dispersion in alignment of different cells was measured by a scan across the Crab Nebula after launch (12 arcmin for LX20, 19 arcmin for LX20, and 17 arcmin for LX30) and by comparing the scan profile with simulations with different amounts of dispersion in collimator alignment. The resulting contours of constant efficiency for 15 keV and 50 keV photons are shown in Fig.~\ref{fig:LAXPCeffic}.

\begin{figure}[h]
\centering
\includegraphics[width=0.54\textwidth,clip]{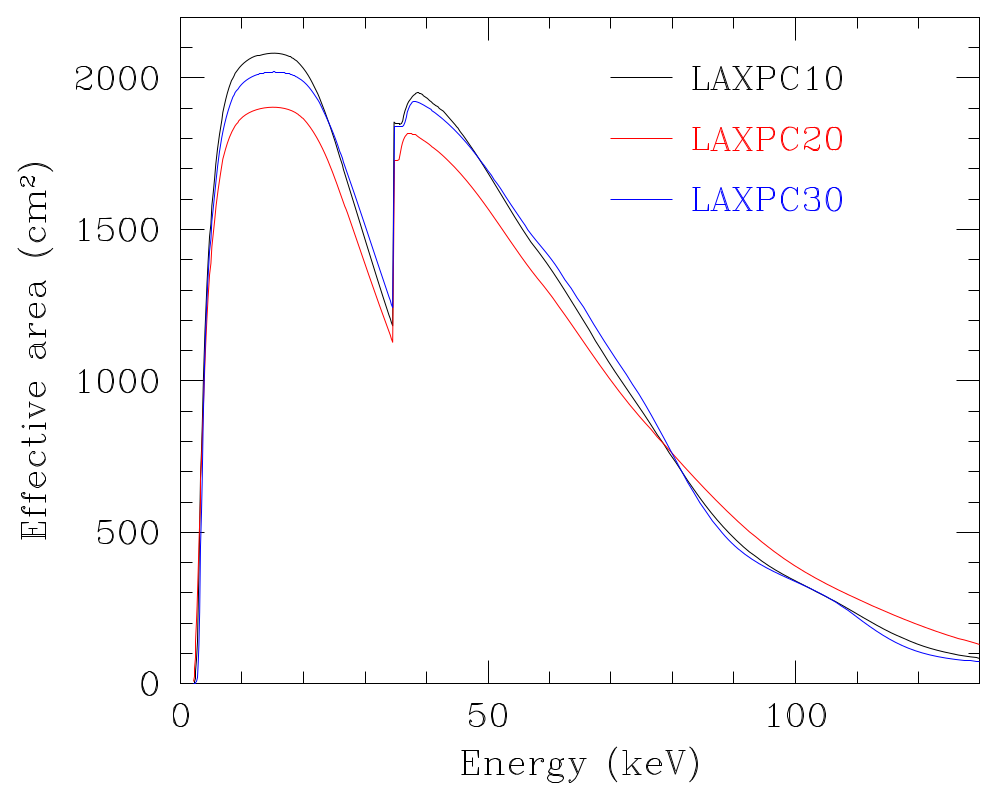}
\includegraphics[width=0.45\textwidth,clip]{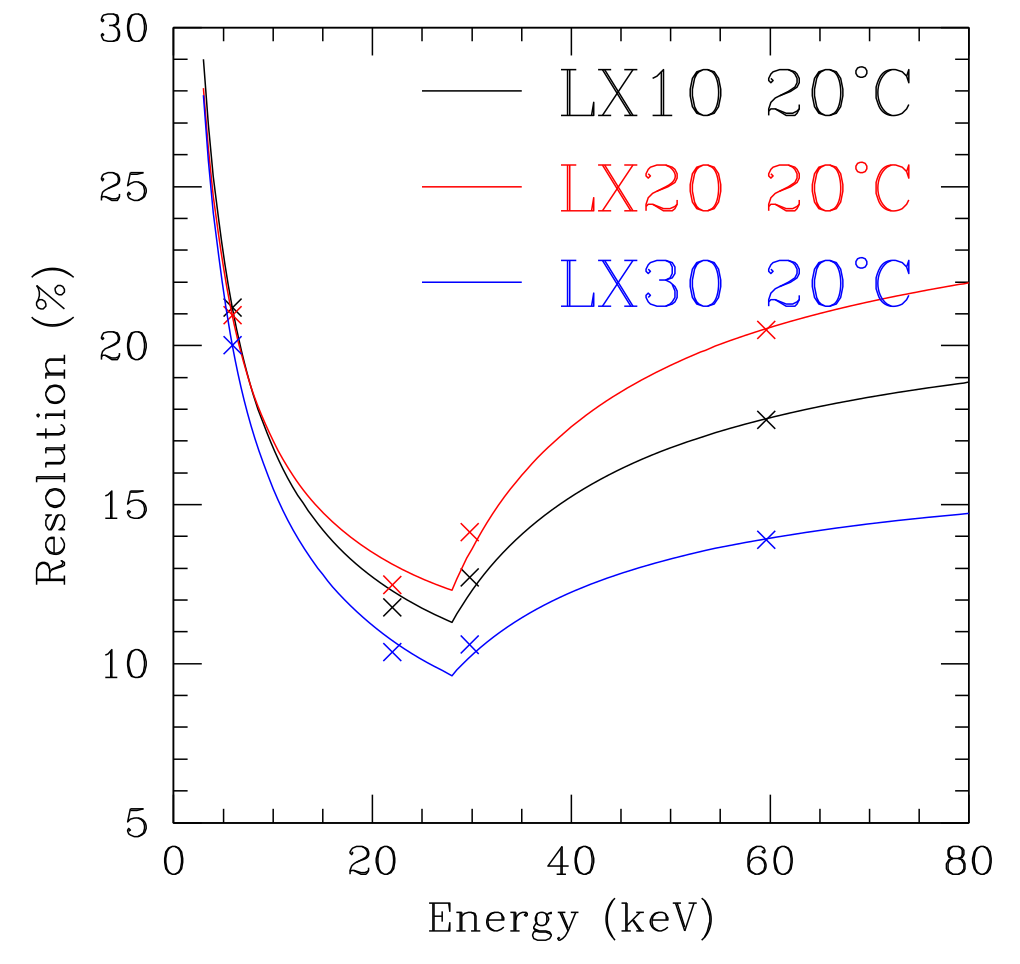}
\caption{Effective areas (left) and energy resolution (right) of LAXPC units. (Credits:~\cite{Antia2017}).}
\label{fig:LAXPCres}       
\end{figure}

The geometric area of 3600 cm$^2$ for each PC reduces to 2800 cm$^2$ due to blockage by the collimator.  The efficiency of detection for photons is a function of energy taking into account the window transmission and absorption in the gas.  This is shown in the left panel of Fig.~\ref{fig:LAXPCres}.  It peaks at $\sim$2000 cm$^2$ between 3 -- 30 keV.  The total effective energy band-width with detection efficiency $>$25$\%$ is 3 -- 80 keV.  The fact that the view axes of the three units are not perfectly aligned with respect to each other, there is an uncertainty of 5$\%$ in the effective areas of the LAXPC units.

Each PC is provided with its own independent front-end electronics (housed at the back of each unit), high voltage supply and signal processing electronics. A system based time generator (STBG),  common for all the three LAXPC detectors, is used to provide a time stamp with accuracy of 10 $\mu$sec for all the accepted events from incident photons.  Outputs of the charge sensitive pre-amplifiers (CSPAs) from 7 X-ray anodes and 3 Veto layer are sent to the peak detectors and events satisfying the selection logic as true X-ray events are further processed by signal processing electronics for (a) Broad Band (BB) counting, and (b) for time tagging of every event to an accuracy of 10 $\mu$s along with its energy.  The identity of the detecting element (detector/anode-layer) is stamped on each event.  

The energy resolution of the detectors and channel to energy mapping were calibrated using three radioactive sources, $^{55}$Fe (5.9 keV), $^{109}$Cd (22 keV/ 24.9 keV) and $^{241}$Am (59.6 keV) on ground and with $^{241}$Am on board. It is shown in right panel of Fig.~\ref{fig:LAXPCres}.  The energy resolution is $\sim$22$\%$ at 6 keV and between 10 -- 20$\%$ from 20 -- 80 keV.  The long term stability of the LAXPC detectors, the peak position and energy resolution of 30 and 60 keV peaks in the veto anode A8 from the on-board radioactive source are monitored regularly, and the adjustments of high voltage and gas purification are carried out at regular intervals as required.  The gain of LX30 had constantly shifted due to a leak after launch, and this unit is no longer operable since 2018.  The gain in LX10 had also been shifting steadily upwards albeit at a rate that is an order of magnitude smaller compared to LX30 and the high voltage for this detector was also adjusted regularly.  LX10 unit is, however, no longer recommended for use since around 1 January 2021.  For the LX20, the gain that had been slowly shifting downwards, and the deterioration in the resolution and gain are both likely due to accumulation of impurities. It had stabilised from 2017 onwards after three gas purifications since the launch.

Each unit of the LAXPC has a large mass and size offering a large cross-section for background particles and also large Compton scattering cross-section in its walls and surrounding materials for photons with energies above 80 keV.  The veto layers are very effective in reducing these backgrounds in the useful energy band of 3 -- 80 keV. However, many background events still get registered thus making these PCs background dominated for weak sources, and thus affecting the sensitivity for source detection.  The total number of background counts averaged over an orbit is $\sim$200 -- 250 counts s$^{-1}$ per PC with a variation of $\sim$20$\%$ around the average value. All PCs are switched off during the passage of the satellite through the South Atlantic Anomaly just south of the equator for safety as the background is very high and can lead to discharge inside the PCs.   The background had changed with time until the middle of 2016 due to changes in gain, but after corrections it decreased a little. The variation in LX10 was larger than that in LX20.   A residual fluctuation at the level of 5 counts s$^{-1}$ in all detectors remains even after all the corrections are applied. This is most probably due to variation in the charged particle flux with time thus limiting the sensitivity. The situation can get worse during the time of geomagnetic disturbances, even minor ones, thereby making the study of sources fainter than $\sim$1 mCrab, irrespective of the observing time, as impossible.
The LAXPC sensitivity allows in ideal conditions, a 1 mCrab source to be detected at 3$\sigma$ level in the 3 -- 100 keV energy band which improves to 6$\sigma$ level for an energy band of 2 -- 10 keV, in about 100 s.

\subsection{\textit{LAXPC Data Analysis}}

The software and responses for analysing LAXPC data are at available at \url{http://www.tifr.res.in/~astrosat_laxpc/LaxpcSoft.html}.  The users should download the L1 data from the astrobrowse site of the ISSDC and use this software on a linux operating platform. The responses and the background are adjusted for variations in gain and resolution, and the database and correction software for this are also made available on the LAXPC site mentioned above.

\begin{figure}[h]
\centering
\includegraphics[width=120mm]{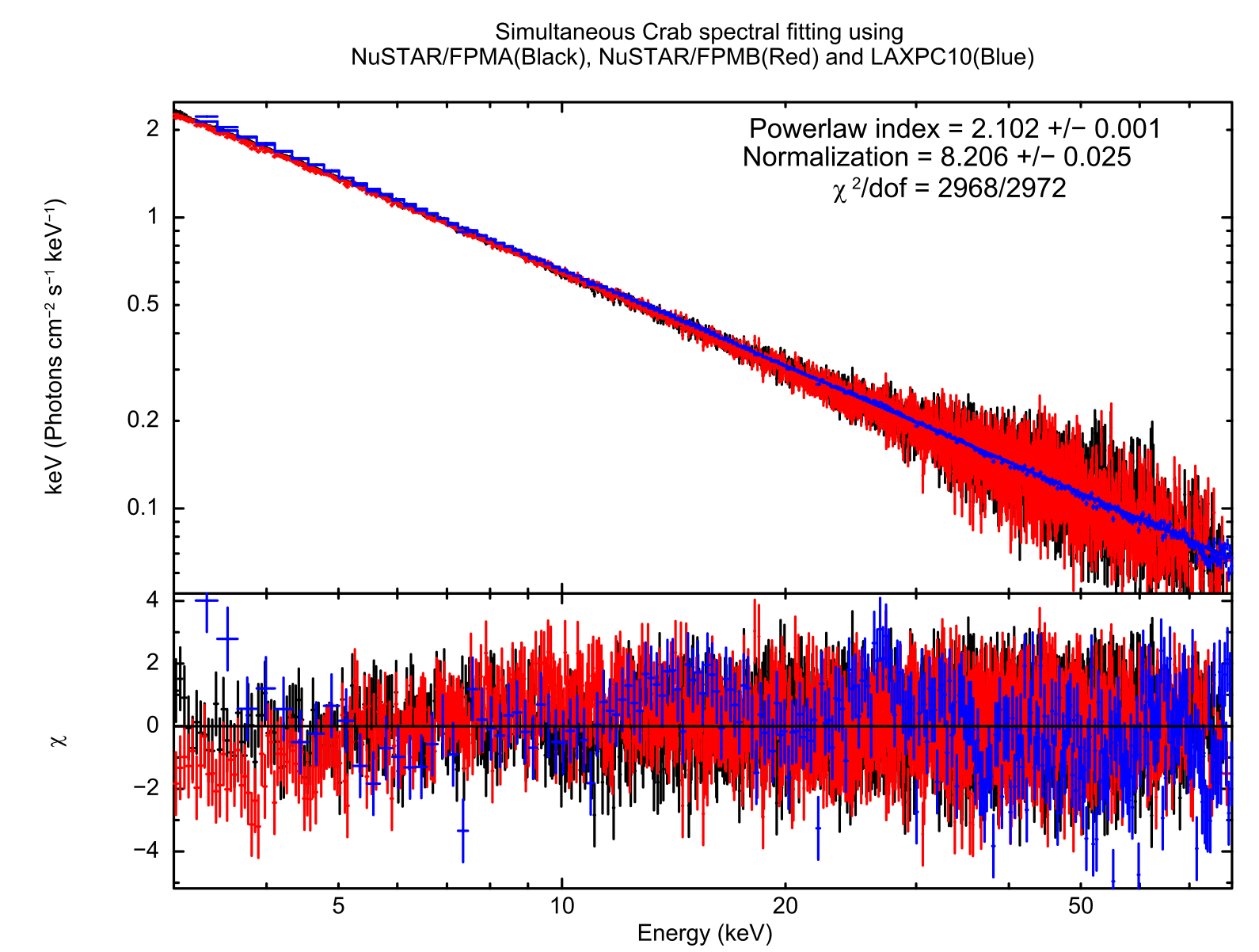}
\caption{A joint fit to NuSTAR and LAXPC10 data from a simultaneous observation on 2016 March 31 of the Crab Nebula. (Credits:~\cite{Antia2017}).}
\label{fig:Crabfit}       
\end{figure}

\begin{figure}[h]
\centering
\includegraphics[width=120mm]{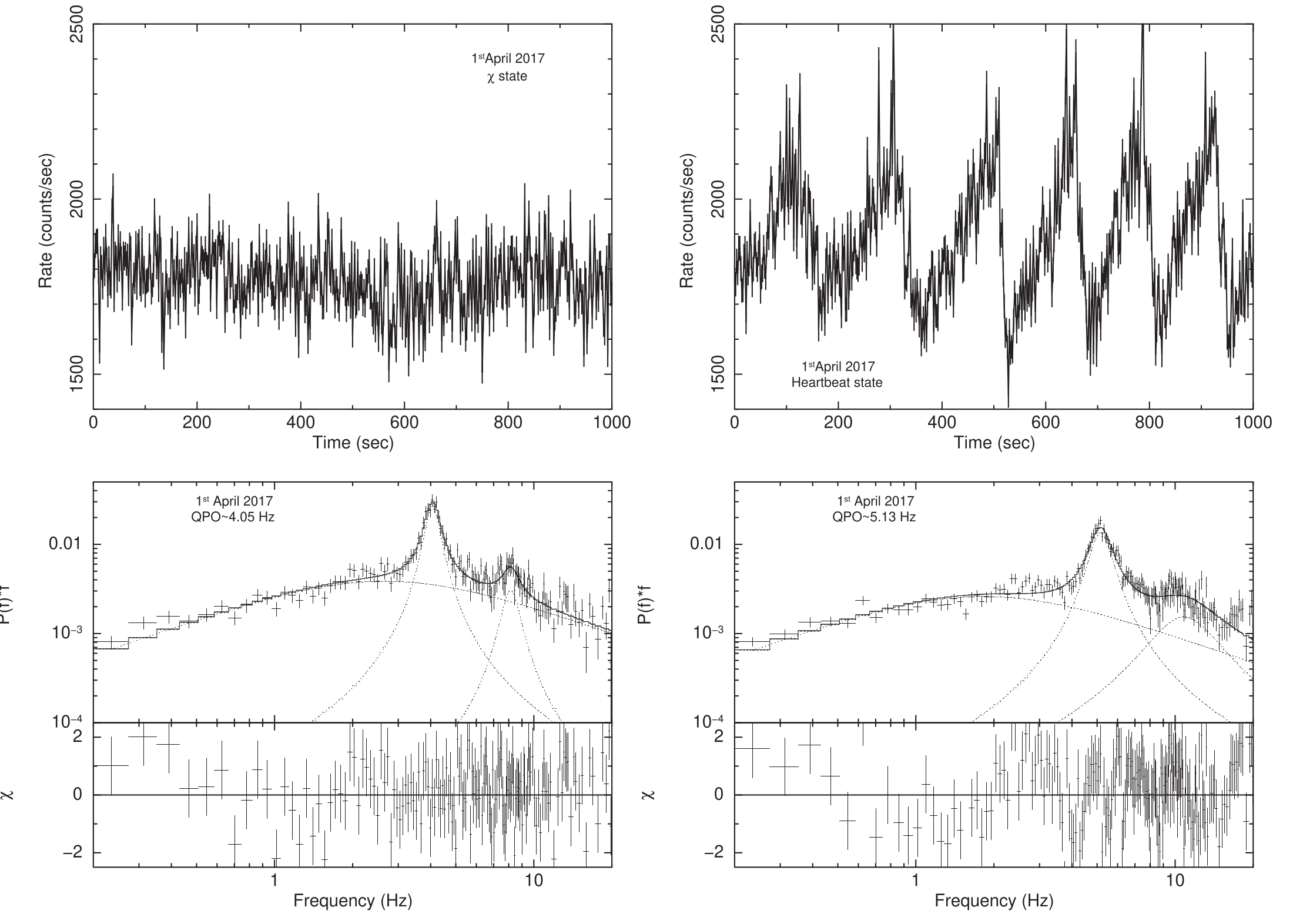}
\caption{X-ray lightcurves in the energy range of 4-50 keV in the $\chi$ class state (top left) and in the heart-beat state (top right) of GRS1915+105 with 2 s bins. The corresponding power density spectra in the 0.2 –- 20.0 Hz range are shown in the (bottom). Credits:~\cite{Misra2020}).}
\label{fig:Timing}       
\end{figure}

An observation of the Crab Nebula was carried out with AstroSat simultaneously with NuStar on 31 March 2016.  A simultaneous fit to data from both the observatories, including 1$\%$ systematics in the LAXPC detector response, gave a reduced $\chi^2$ of 1.0 and is shown in Fig.~\ref{fig:Crabfit}.  The relative normalisations with respect to NuSTAR of the three LAXPC detectors were calculated to be 0.92, 0.84, and 0.89, respectively for LX10, LX20, and LX30.   An example of the timing capability of the LAXPC is shown In Fig.~\ref{fig:Timing}.
	
For more details of the LAXPC and its performance in the first 5 years after launch see \cite{Antia2017}, \cite{Antia2021}) and the handbook posted at \url{http://astrosat-ssc.iucaa.in/documents}.

\section{\textit{Soft X-ray focusing Telescope (SXT)}}

\begin{figure}[h]
\centering
\includegraphics[width=115mm]{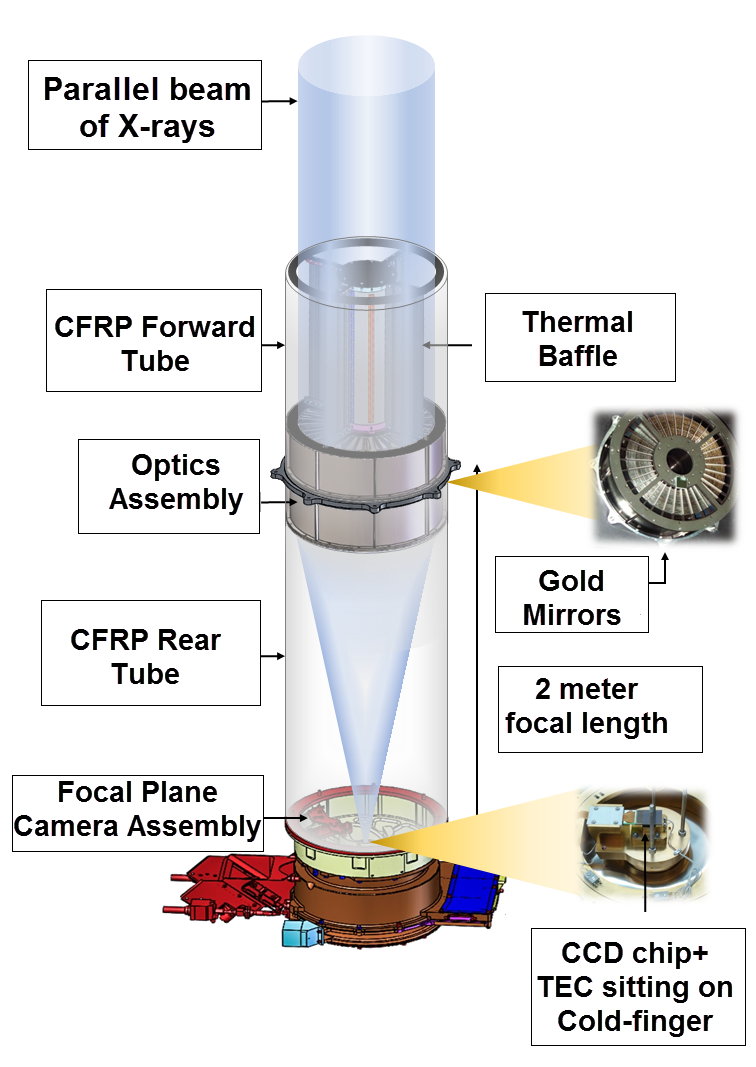}
\caption{A schematic of the SXT showing its major components. Credits: H. Shah of the SXT team.}
\label{fig:SXT}       
\end{figure}

The SXT consists of a grazing incidence telescope with a focal plane camera carrying a cooled charge coupled device (CCD).  It focuses soft X-rays on to the CCD for  medium resolution spectral studies and low resolution temporal studies in 0.3 -– 7.0 keV region.  The optical system with the camera weighs 52 kg and has an envelope of 0.382m x 0.580m x 2.482 m length.  The electronics box kept within 1.5 m of the CCD pre-amplifier weighs additional 13 Kg.   Major components of the SXT are shown in a schematic diagram (Fig.~\ref{fig:SXT}). These are: (i) A mirror assembly based on conical approximation to Wolter I type telescope, as in the Japanese X-ray telescope used in ASCA and Suzaku X-ray observatories.  The mirror assembly has a 1$\alpha$ section (approximate paraboloid) and a 3$\alpha$ section (approximate hyperboloid); (ii) A Focal Plane Camera Assembly (FPCA) housing a cooled CCD; (iii) A deployable cover/door at the top end of the telescope that covered the optical elements on the ground and protected them from contamination.  This was opened permanently $\sim$2 weeks after launch and is perched at an angle of 256 degree with respect to the +roll axis; (iv) A “Thermal Baffle” placed between the mirror assembly and the telescope door made of anodised aluminum alloy 6061 T6. It protects the telescope from the Sun, and  provides a base for mounting the heaters to maintain the optics within a certain specified range of temperatures, and to block the unwanted area of the optics. The sun avoidance angle with the thermal baffle is $\sim$45 degree; (v) A “Forward tube” made up of Composite Fibre Reinforcement Plastic (CFRP) extending from the bottom of the top cover, covering the thermal baffle assembly as well as the 1$\alpha$ section of the optics assembly; (vi) A metallic ring (Ring 1) provides an interface between the Rear tube 1 and the middle flange of the optics from the bottom side. Another ring (Ring 2) provides an interface between the Forward tube and the Rear tube 1 from the top size; (vii) A Rear tube-1 also made up of CFRP is a hollow cylinder of diameter 343 mm ID and 347.8 mm OD and extends from Ring-1 to the “Deck Interface Ring” (DIR). It houses 3$\alpha$ optics while the Forward tube houses the 1$\alpha$ assembly; (viii)  A Deck Interface Ring” (DIR) made of Al alloy 6061 is used to assemble rear tube-1 and rear tube-2 (shown together in the Fig. 16), and to provide interface between the payload and the top deck of the satellite.  The rear tube-2 made of CFRP is a hollow stepped cylinder with a top portion thicker than the bottom portion to provide stiffness, and it extends from the DIR to the FPCA interface ring.

X-rays are incident at grazing angles, and are first reflected by the 1$\alpha$ section, and then reflected by the 3$\alpha$ section to the CCD at the prime focus of the telescope.  Mirrors are made of very thin (0.2 mm thickness) Al foils covered with replicated gold surfaces using the same technique as employed in the Suzaku. Replication of gold surfaces was carrried out using glass mandrels leading to surfaces with a roughness of $\sim$1 nm, thus producing mirrors with low scattering of X-rays.  Each mirror was made for a quadrant of a shell thus four of them make one shell of 1$\alpha$ section, and four more make the corresponding co-axial shell of the 3$\alpha$ section. Thus,  eight mirrors make up one complete shell of the grazing incidence telescope.  The SXT has 40 complete shells nested within each other to increase the reflecting area for X-rays and, therefore, consists of a total of 320 mirrors. Each mirror is 100 mm long with the radius of the outermost and innermost shell being 130 mm and 65 mm, respectively with the other mirrors having radii lying in between these values as determined by ray-tracing to provide a clear path for the X-rays after reflection. A specially machined holder was made to house each mirror so that all the mirrors in all the shells were mounted co-axially. The overall performance of mounted mirrors was measured on ground using optical light -- targeting each foil with a pencil beam for adjustments followed by a broad parallel beam from an inverse optical telescope for the overall performance (see \cite{Navalkar2021} for more details).

\begin{figure}[h]
\centering
\includegraphics[width=115mm]{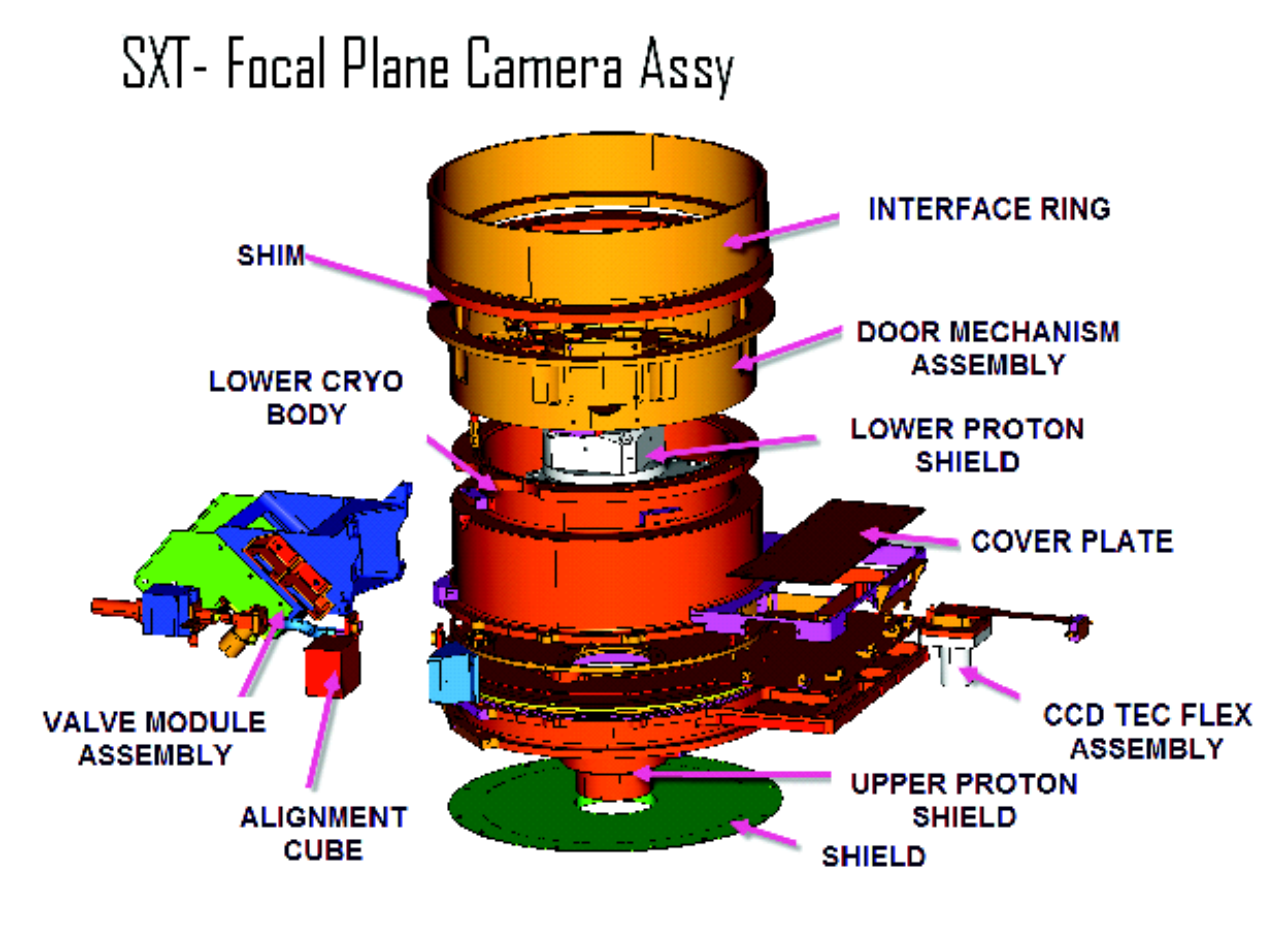}
\caption{A schematic of the FPCA showing its major components. Credits: \cite{Singh2014}}
\label{fig:FPCA}       
\end{figure}

The X-ray sensor/detector in the SXT is a charge coupled device (CCD) in the FPCA, located at a distance of 2 m measured from the middle of the two conic sections.  Fig.~\ref{fig:FPCA} shows the FPCA in the SXT. Photons from a source reach the CCD through a very thin and fragile window made of a polyimide film (184 nm thick and coated with 48 nm of Aluminum on one side) which blocks the ultra-violet and visible photons but allow the X-rays to enter. The optical transmission of the filter is less than 5 10$^{-3}$ (similar to the thin filter used in the EPIC-MOS cameras in the XMM-Newton observatory).  The filter design provides $\sim$7 magnitude of optical extinction over the visible band. The CCD is surrounded by a proton shield to block the charged particles from all other sides. CCD-22 -- a MOS device built by the e2V Technologies Inc., UK, for the European Photon Imaging Camera (EPIC) aboard the XMM-Newton and Swift observatories and supplied to the University of Leicester, is used as the detector in the FPCA.  It is a three-phase, frame transfer CCD with open electrode  structure for a useful band pass of 0.2 to 10 keV  with imaging area of 610 x 602 array of 40 x 40 micron pixels including an over-scan and a storage region of 600 x 602 array of 38 x 12 micron pitch. The image pixel size corresponds to $\sim$4 x 4 arcsec$^2$ on the sky.   The readout and pre-amplifier is part of the FPCA and the voltages and signals are communicated using a flex-cable.  The CCD in focal plane camera is operated at -82 degree C using a thermo-electric (TEC) cooler connected to a radiator plate on the cold side of the satellite via an ethane heat pipe.  The radiator plate gives a maximum temperature of -40 degree C at the junction between the heat pipe and the camera cryostat.   The performance of the CCD and its associated processing electronics is monitored in real time using radioactive fluorescent sources at the 4 corners of the detector. Another calibration source mounted on the inside of the camera door was used for ground tests and during the flight until the door was opened permanently on 26 October 2015.  

The main processing electronics onboard consists of ten cards of circuits and uses three Field Programmable Gate Arrays (FPGAs).  It is designed to read out the CCD in several modes, the most commonly used one being the photon counting (PC) mode with a time resolution of 2.3775 s, and a faster mode (Fast Windowed Photon Counting  or FW) provides a time resolution of 0.278s for the central 150 x 150 pixel$^2$ window. Other modes are: "Bias Map" (BM) mode, "Calibration" mode (CM), and "House Keeping" (HK) mode for health parameters of the electronics system. Events above a specified threshold energy of 105 eV which is 4$\sigma$ above the noise peak  (it can be set between 100 -- 200 eV through a tele-command by the SXT team) only are recorded.  It is, however, recommended that low energy threshold for analysis of SXT data should be set above 300 -- 350 eV (above the Carbon K-edge of the window).   The energy information for each photon is made available in both PC and FW modes used for science.

\begin{figure}[h]
\centering
\includegraphics[width=120mm]{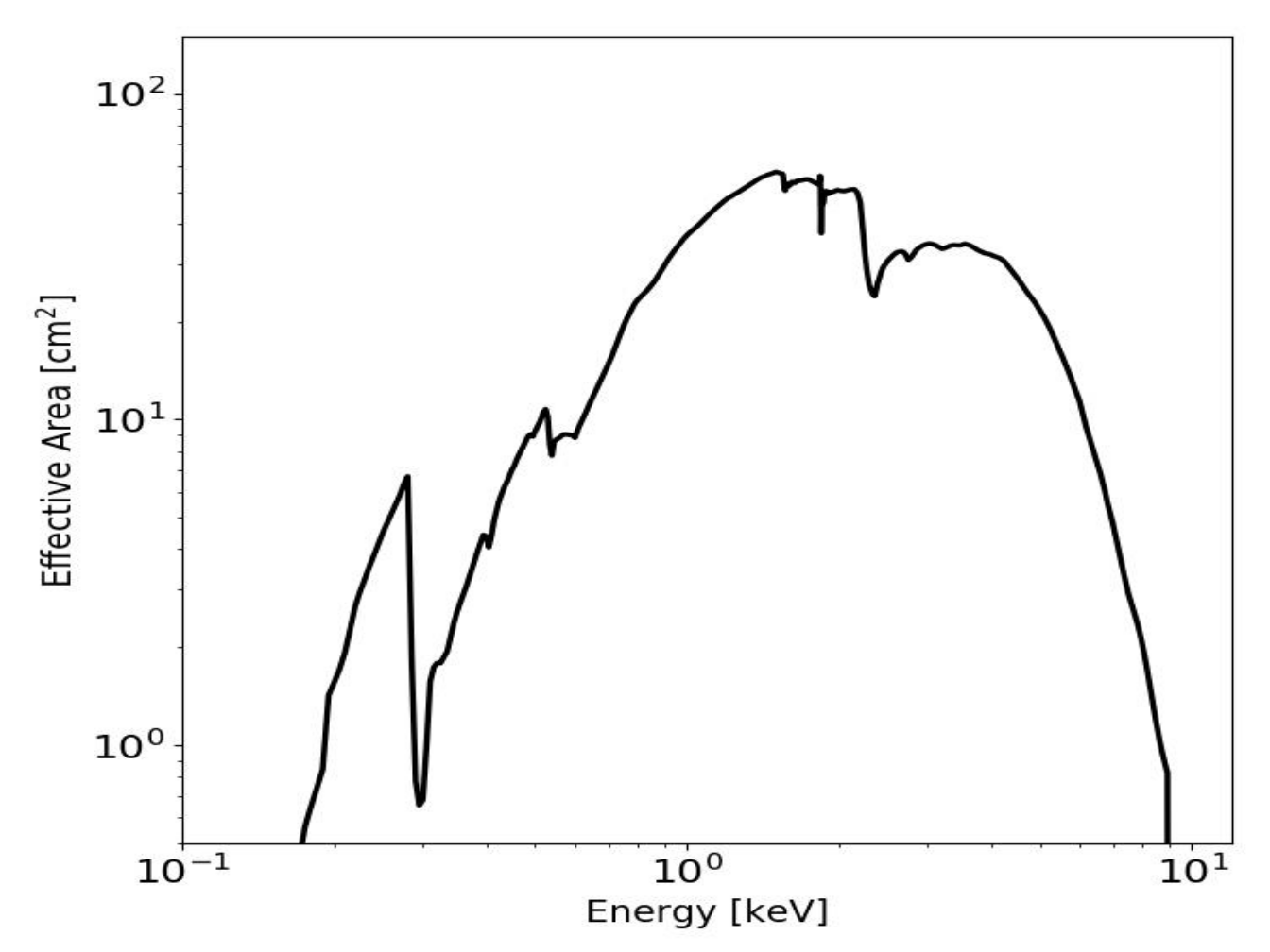}
\caption{Effective area of the SXT (see text for more details). Credits: SXT team at TIFR.}
\label{fig:SXTeffic}       
\end{figure}

The effective area of the SXT as a function of energy that includes the quantum efficiency of the CCD for (isolated -- grade 0 and bi-pixel events grades 1-4), and absorption by the optical blocking filter as measured for the photons incident on-axis is shown in Fig.~\ref{fig:SXTeffic}.

\begin{figure}[h]
\centering
\includegraphics[width=120mm]{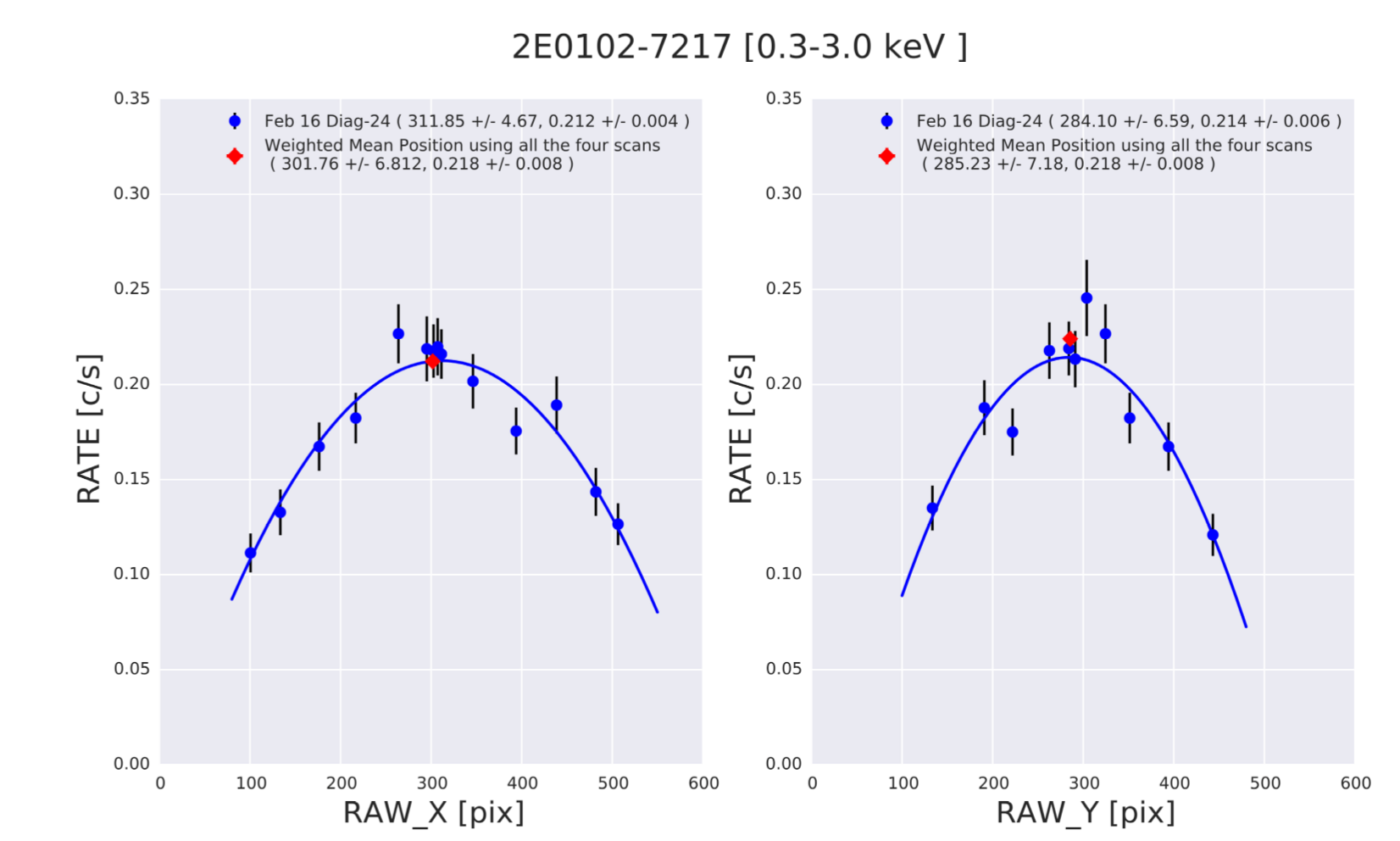}
\caption{A scan across the SNR 2E0102-7217 to determine the boresight of the SXT. Credits: \cite{Singh2017a}.}
\label{fig:snr}       
\end{figure}

The SXT has a field of view of $\sim$40 arcmin. Its spatial resolution was measured in-flight by observing point cosmic X-ray sources.  Observations of 1E0102.2-7219, a supernova remnant in the Small Magellanic Cloud were used by pointing it on different parts of the CCD to determine the boresight of the telescope. This source emits mostly in soft X-rays in the energy range of 0.3 -- 3 keV only.  The results from one of the scans are shown in Fig.~\ref{fig:snr}. The observations led to the determination of the bore sight of the SXT close to centre of the FoV at X = 302($\pm$7) and Y = 285($\pm$7) pixels.  These observations also provided the vignetting in the telescope area as a function of off-axis angle. 

\begin{figure}[h]
\centering
\includegraphics[width=0.49\textwidth,clip]{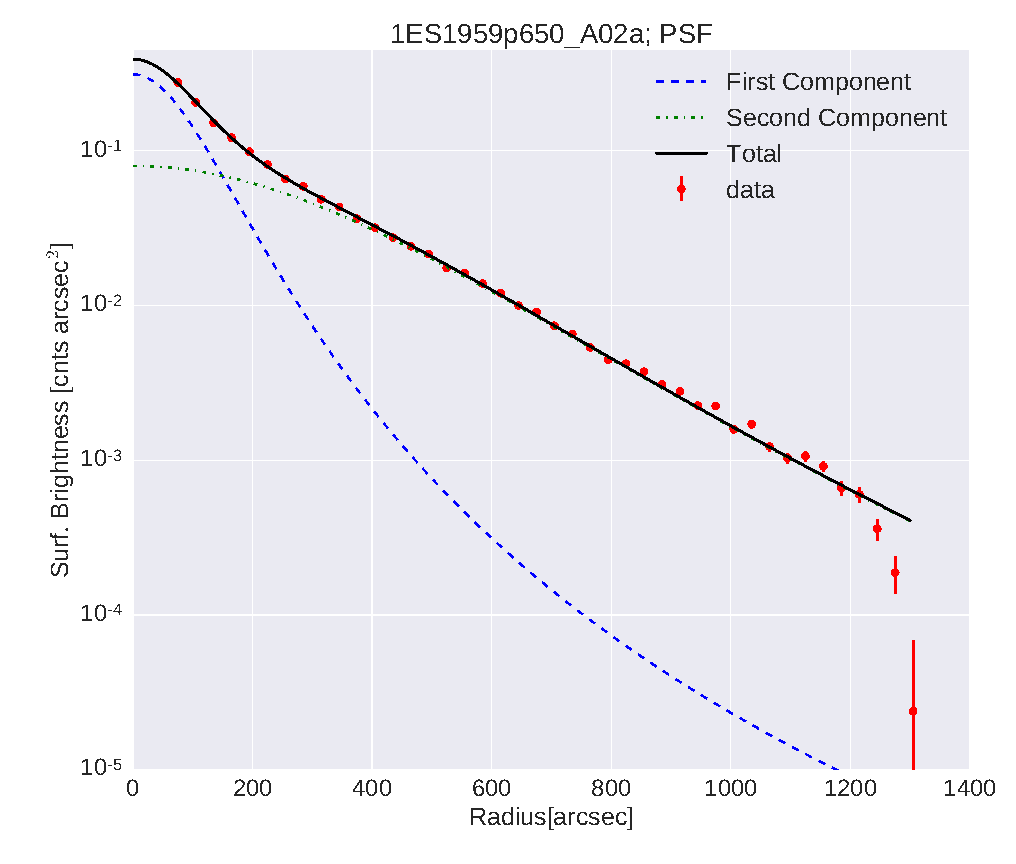}
\includegraphics[width=0.49\textwidth,clip]{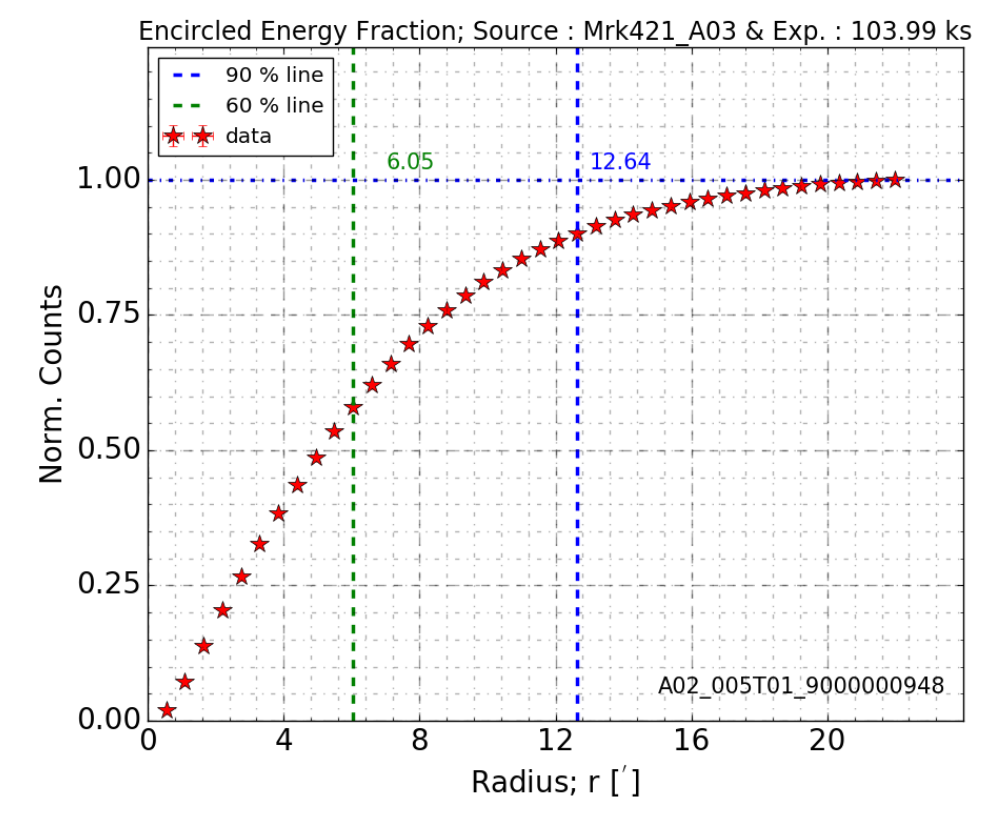}
\caption{Point spread function of the SXT. Credits: S. Chandra of the SXT POC team.}
\label{fig:SXTPSF}       
\end{figure}

The point spread function (PSF) of the SXT has been derived from several observations of many point sources.  In Fig.~\ref{fig:SXTPSF} an example of the PSF and EEF (Encircled Energy Fraction) derived from observations of 1ES1959+65 and Mrk 421 blazars for the energy range 0.3 -- 7.0 keV are shown. These data have been fitted with double King profiles. The central parts of the PSF are well fitted by the core radius given by the central King function, which is determined to have a mean value of 76 $\pm$ 24 arcsec.  If expressed in terms of the full width half maximum (FWHM), the PSF is $\sim$150 arcsec. The second King model, indicative of the scattering in the mirrors and their relative alignments,  has a peak brightness of 7$\%$ of the central King profile but core radius of $\sim$570 $\pm$ 28 arcsec.  The EEF values indicate that to get the full intensity of a source we need to use an extraction radius of between 12 -- 16 arcmins, depending on the intensity of the source. It also implies that the background has to be taken from deep exposures of blank sky regions.

\begin{figure}[h]
\centering
\includegraphics[width=120mm]{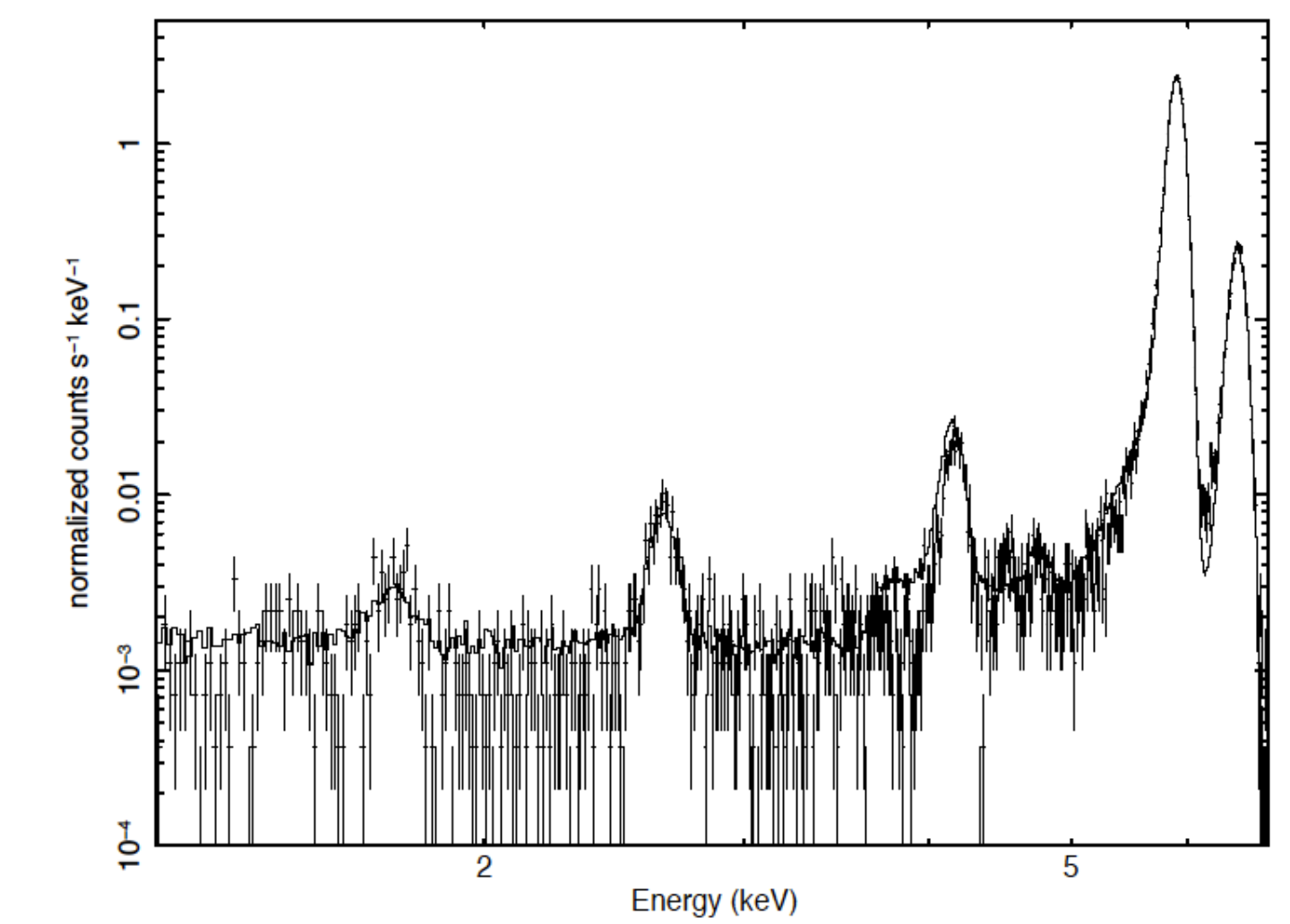}
\caption{An X-ray spectrum obtained from four $^{55}$Fe calibration sources illuminating the four corners of the CCD (size 80 x 80 pixels) illustrating the energy resolution obtained. The two prominent lines are due to Mn K$\alpha$ (5.895 keV) and Mn K$\beta$ (6.49 keV).  Other lines are due to the escape of photons from these two lines and the flourescence of Al, Si,  Cl and Ti. The response function has been generated based on these data and a fit of the response model to the data is shown as a histogram. Credits: \cite{Singh2016}}
\label{fig:SXTcal}       
\end{figure}

The energy resolution as measured using the calibration sources is shown in Fig.~\ref{fig:SXTcal}.  Using even longer exposures than shown here, it is determined to be $\sim$136 eV (2.3$\%$) at 5.9 keV and ~6$\%$ at 1.5 keV. The background in a CCD frame is extremely low ($\sim$0.2 counts s$^{-1}$) over the entire FoV in the energy band of 0.3 -- 7.0 keV. Thus, focusing on a detector with a very small size and mass provides good sensitivity particularly for sources with X-ray spectrum between 0.3 -- 2.0 keV where the effective area of the SXT is highest.  The natural ability of a pixelated detector such as a CCD to distinguish between patterns of charge deposited by X-rays and particles helps to further reduce the background.  For a typical detection radius of 10 arcmin on the CCD for a weak source, the background is $\sim$0.07 counts s$^{-1}$ . Thus in a 10000 s exposure, the 5$\sigma$ point source detection limit is $\sim$0.015 counts s$^{-1}$ above the background, which roughly corresponds to 6 10$^{-13}$ ergs cm$^{-2}$ s$^{-1}$ ($\sim$15 $\mu$Crab).  Considering that the SXT has a peak effective area of only $\sim$80 cm$^2$, it provides nearly two orders of magnitude better sensitivity than a detector of similar area but without any focusing.  

\subsection{\textit{SXT Data Analysis}}

The raw data obtained from the satellite are processed and made available as L1 data by ISRO to the SXT POC at TIFR, Mumbai, for verification and processing through the SXTPIEPLINE after which both L1 and L2 data are posted at the “Science Data Archive for AstroSat Mission” accessible from \url{https://astrobrowse.issdc.gov.in/astro_archive/archive/Home.jsp}. 
Conversion to the L2 data uses calibration data files from SXT Calibration database, CALDB as input.   The processing involves, event extraction using SXTEVTGEN, time tagging of events using SXTTIMETAG, coordinate transformation from raw to detector and XY co-ordinates using SXTCOORD, bias subtraction and adjustment using SXTBIASSUB and SXTBIASADJ, flagging of bad pixels and calibration source events using SXTFLAGPIX, events grading and PHA construction for each event using SXTEVTGRADE, search for hot and flicker pixels using SXTHOTPIX and then carrying out PHA to PI conversion of events using SXTCALCIPI.  
The data processing thus generates an unfiltered event file for grades 0 -- 12 (single pixel – grade 0 or bi-pixel grades 0-4 can further be selected by filtering at a later stage using XSELECT for the analysis of L2 data).  The tool SXTFILTER is used to create L2 MKF file. To generate cleaned event file, data screening is done using tool SXTSCREEN on unfiltered event file utilizing L2 MKF file, HK and event range files from calibration database.  Basic L2 data products such as image, light curve and spectrum are generated utilizing clean event file by product generation tool called SXTPRODUCTS which has been designed based on XSELECT interface from HEASARC (NASA). The L2 data products so generated are further processed and corrected for proper exposures. The target source centroid position is determined with associated errors using SXTCENTROID.  The SXTMKARF generates corrected ARF accounting for vignetting, psf and exposure correction based on exposure map. The exposure map is generated using SXTEXPOMAP, which accounts for the loss of flux due to marked bad pixels and columns.   Higher-level data files are compatible for use with standard astronomical tools such as XRONOS, XIMAGE, XSPEC and FTOOLS etc.  
Instructions for analysing SXT L2 data in order to extract science can be found at \url{http://astrosat-ssc.iucaa.in/uploads/sxt/Analysing_SXT.txt} and at \url{https://www.tifr.res.in/~astrosat_sxt/index.html}. 
The required response matrices (RMF and ARF) and background files can be found at \url{http://astrosat-ssc.iucaa.in/?q=data_and_analysis} 
and at \url{https://www.tifr.res.in/~astrosat_sxt/index.html}. 

The telescope response matrix (ARF file) requires that the extraction radius be in the range of 12 - 16 arcmins for obtaining proper normalised flux values. For a smaller extraction radius users should apply a correction tool to the default ARF by running the utility 
sxt\_ARFModule\_v02.py downloadable from the above sites. 
Comparisons with the other CCD based observatories such as, XMM-Newton, Swift XRT etc. using the IACHEC models and other observations show that the cross normalisation values obtained from the SXT are accurate to within 10$\%$ over the energy range of 0.3 -- 7.0 keV (also see below).  A nominal gain shift (or offset) of about +20 to +50 eV in the spectral response is sometimes required to improve the spectral fitting of data, and it can be judged by examining the residuals around a sharp absorption edge near 2.1 -- 2.2 keV due to the M-edge of the gold used on the surface of the mirrors. It is recommended that the SXT data be selected to lie within the energy range of 0.35 and 7.10 keV for all analyses. 

All users are encouraged to download and use only the L2 data for analysis. The L2 data cleaned events files are made available for each orbit.  These event files should then be merged by users into a single cleaned event file using a Julia or python based merger tool available from the above sites to avoid time-overlapping events from the consecutive orbits which can creep in sometimes.
 
Note that there is a small but significant offset $\sim$a few arcmins between the SXT pointing and the pointing of LAXPC, CZTI and UVIT.  Therefore, the proposers should use the PC mode when SXT is not the primary instrument. This is because, in the case of the FW mode the source may not be fully covered within the SXT foV. However, one may need to use the SXT FW mode for some science goals in order to reduce pile-up and/or to have better time resolution. In such a case, proposers should make SXT the primary instrument, even if SXT does not serve the primary science.

\begin{figure}[h]
\centering
\includegraphics[width=0.52\textwidth,clip]{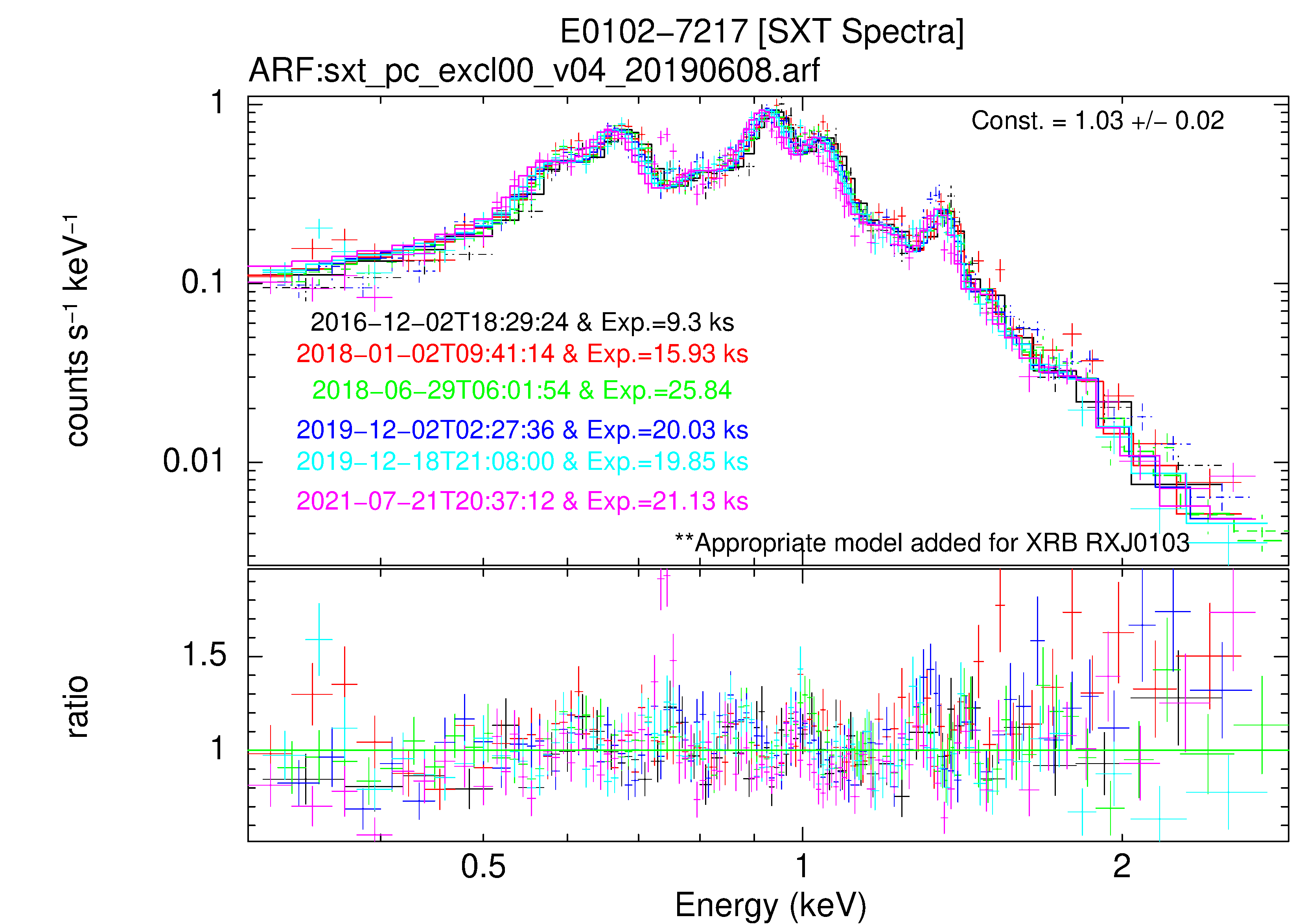}
\includegraphics[width=0.47\textwidth,clip]{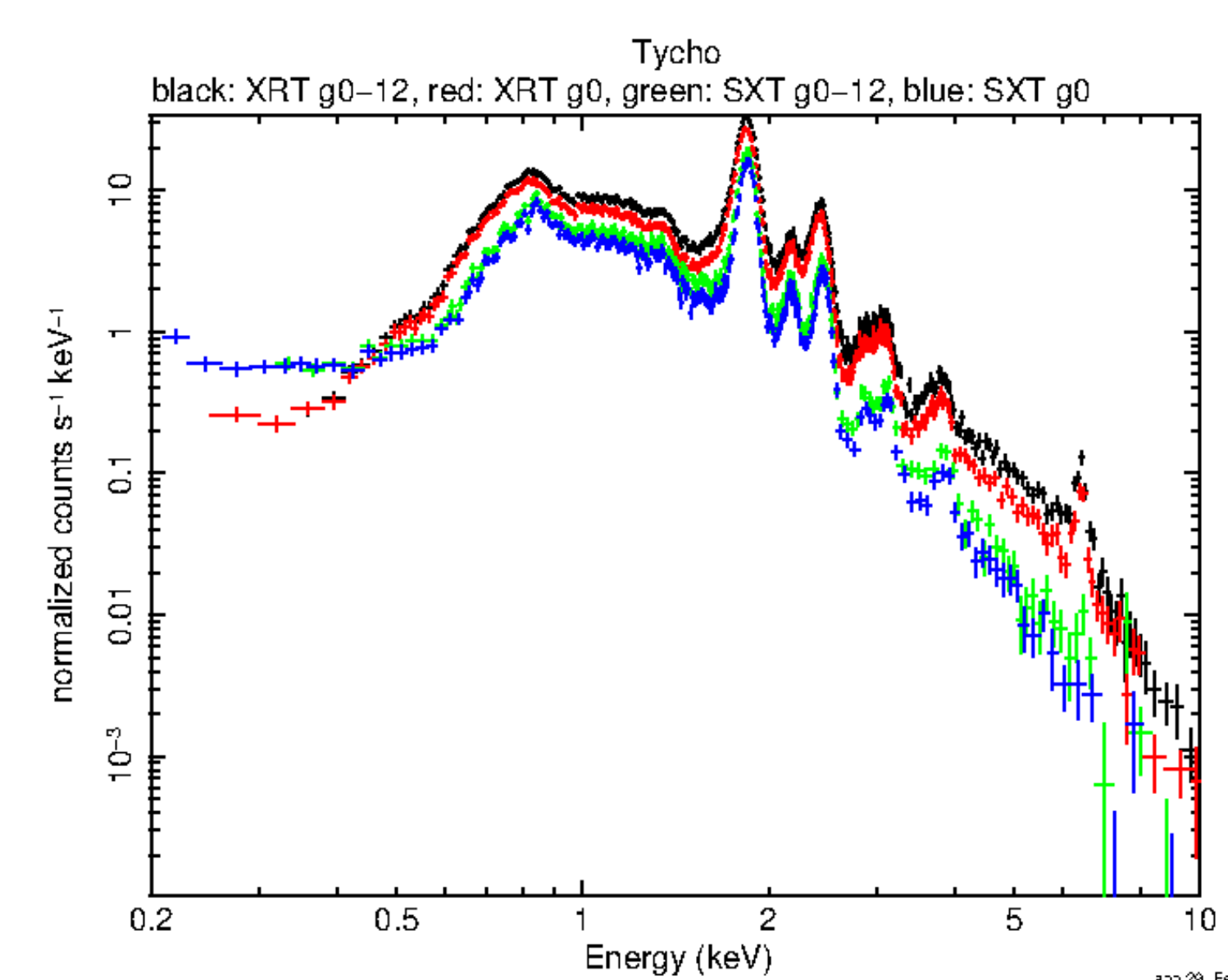}
\caption{IACHEC Model for 1E0102-7217 fitted to the SXT spectra taken regularly over the last 5 years to monitor the low energy calibration with respect to several instruments like Chandra ACIS, XMM-Newton MOS etc is shown on the left. A comparison of the AstroSat-SXT and Swift XRT spectra of Tycho SNR on the right. Credits: S. Chandra and \cite{Singh2017a}.}
\label{fig:SXTIACHEC}       
\end{figure}

Spectral calibration of the SXT was carried out by observing the supernova remnants: Tycho, Cas A and 1E 0102-7217. The spectra of 1E0102 are monitored annually, and have been fitted with a common spectral model known as the International Astronomical Consortium for High Energy Calibration (IACHEC) standard, empirical model published in \cite{Plucinsky2017} and available at \url{https://wikis.mit.edu/confluence/display/iachec/Thermal+SNR}. The results of the fit to the SXT data from 2016-2021 are shown in Fig.~\ref{fig:SXTIACHEC} (left). These results show that with the currently available response the agreement with the model norm is within 5$\%$ in this soft energy band.  An older comparison of the SXT spectra of Tycho taken in 2016 with the spectra obtained from the Swift X-ray observatory is also shown here in Fig.~\ref{fig:SXTIACHEC} (right) for event grades 0 and grades 0-12.

Further details about the SXT can be found in \cite{Singh2014}, \cite{Singh2016}, \cite{Singh2017a}, \cite{Singh2017b} and references therein.

\section{\textit{Cadmium–Zinc–Telluride Imager (CZTI)}}

\begin{figure}[h]
\centering
\includegraphics[width=120mm]{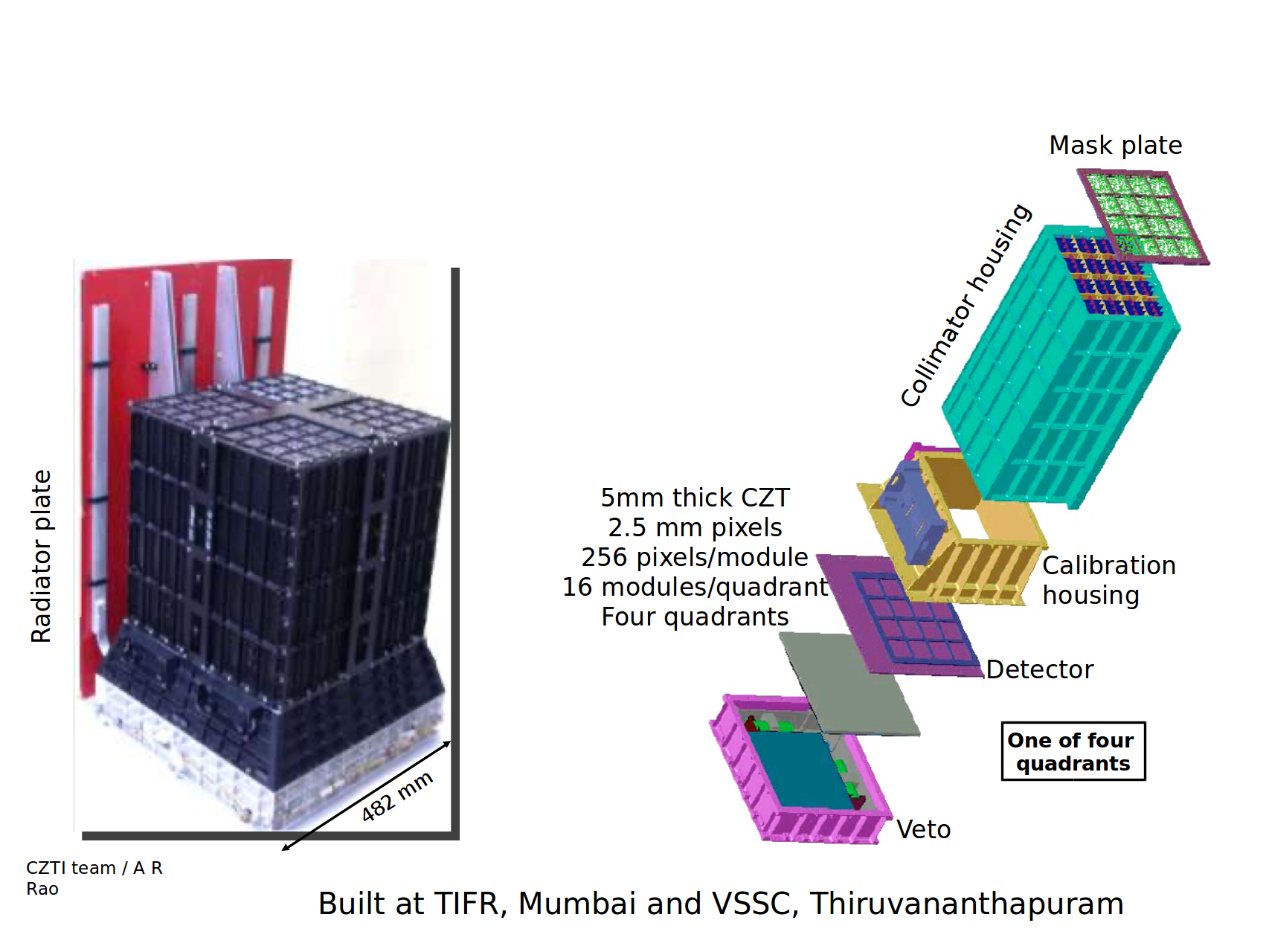}
\caption{An image of the CZTI (left) and a schematic of one of its quadrants (right). Credits: CZTI team}
\label{fig:CZTI}       
\end{figure}

This instrument was built to detect very hard X-rays in the 15 -- 200 keV energy range.  It weighs 49.5 kg and has an envelope of 0.53 m (length) x 0.51 m (width) x 0.813 m (height).  The CZTI consists of 64 detector modules made of 5 mm thick crystals of compound semiconductor Cd$_{0.9}$Zn$_{0.1}$Te$_{1.0}$ (CZT). These 64 modules are arranged in four identical and independent quadrants. A continuous anode made of 50$\mu$m thick aluminized mylar runs across each module. The cathode is divided into a grid of 16 x 16 pixels which are directly bonded to two Application Specific Integrated Circuits (ASICs) for readout. The ASIC signals are further read and processed by the main electronics unit. The modules are arranged in four quadrants with 16 of them in each quadrant. Scintillation detectors, known as Veto Detectors, made of CsI(Tl) mounted on photomultiplier tubes operated at 800 V,  are located below the CZT modules and serve as active anti-coincidence shields to reject charged particle events and very high energy photons.  A radioactive calibration source, $^{241}$Am, is accommodated in a gap of  $\sim$8 cm between the base of the collimator slats and the detector plane, in each quadrant. These sources shine alpha-tagged 60 keV photons on the CZT detectors.  All detectors are connected to a passively cooled radiator plate located on a side of the satellite facing the satellite +Yaw axis.  The CZT detectors are thus maintained at a temperature of $\sim$0 degree C to reduce the thermal noise. X-ray events detected by CZTI have a time resolution of 20$\mu$s. The imaging capability is provided by a coded aperture mask located 481 mm above the detector plane giving 17 arcmin resolution over a 4.6 x 4.6 degree FoV. The mask for each quadrant is made of a 0.5 mm thick tantalum plate with square and rectangular holes matching the pitch of the CZT detector pixels. These holes cast a unique shadow on the detector plane for each source direction. The patterns are based on 255-element pseudo-noise Hadamard set of uniformly redundant arrays.   A schematic diagram of the CZTI is shown in Fig.~\ref{fig:CZTI}.

\begin{figure}[h]
\centering
\includegraphics[width=120mm]{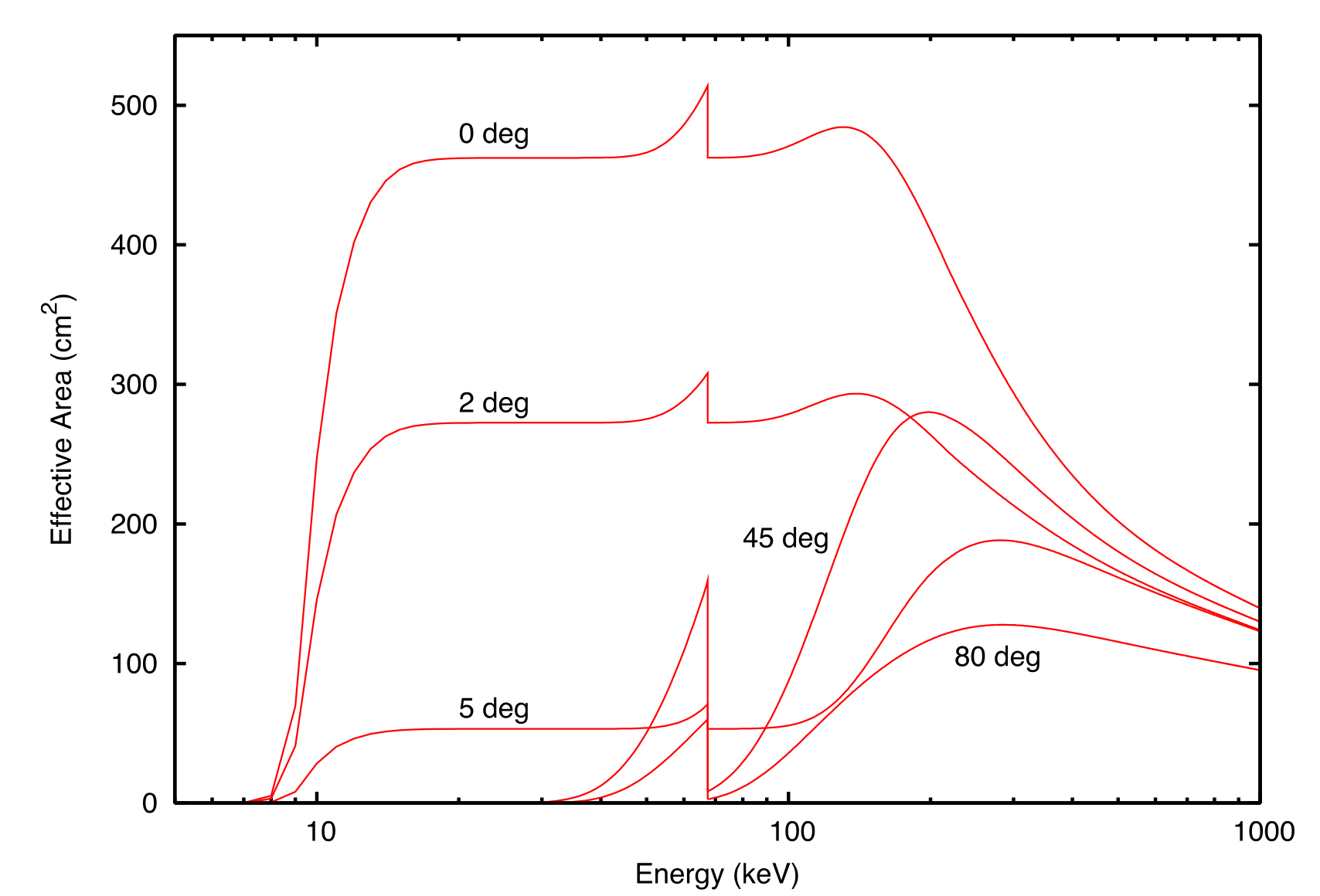}
\caption{Effective area as a function of energy of the AstroSat CZT Imager at normal incidence (0 deg) and several other off-axis angles (Credits: CZTI Team)}
\label{fig:CZTIeffic}       
\end{figure}

The energy bandwidth of the CZT detectors is 15 -- 200 keV.  The effective area of the CZT Imager as a function of photon energy and illumination angle is shown in Fig.~\ref{fig:CZTIeffic}.  It has been estimated by accounting for the energy-dependent transmission through the various surfaces making up the CZTI structure, and the energy-dependent absorption by the CZT detector, at different angles of incidence. The peak near 67 keV occurs due to K-escape of tantalum.

Other main components of the CZTI are: the Front-end Electronics Board (FEB) communicating with the CZT modules and a Processing Electronics (PE) box.  An FPGA is used in FEB to handle the signals from the three different detectors (CZT, Alpha and CsI Veto). Data are stored in a memory and transmitted to PE, every second. The coincidence among different detectors are performed digitally by ground data processing. The PE,  housed separately,  contains an FPGA with an embedded processor and controls all satellite interfaces, detector interfaces and handles onboard memory and data management. The functions of the PE include reading, analysing, storing and/or transferring detector data to satellite via data formatter. The PE also controls the FPGA in the FEB using 16-bit serial commands.

The CZTI has an unobstructed view of the sky for energies $>$100 keV, and is therefore sensitive to Gamma Ray Bursts and other bright transients occurring anywhere in 30$\%$ of the sky around the observing axis of the satellite.  Cosmic diffuse $\gamma$-rays and $\gamma$-rays originating from the satellite structure (spallation background) due to the interaction of cosmic rays are the primary sources of background in the CZTI. To distinguish these background counts from the real source counts a slab of CsI is placed under each block of CZT.   High energy photons which deposit energy in CsI as well as CZT are rejected by anti coincidence techniques and hence do not contribute to the background counts in CZT. Only the photons which deposit partial energy in CZT but do not interact in the CsI block generate background counts in the (10 -- 100) keV region. Background count estimates due to single Compton scattering have been simulated by summing up all the contributions over the entire energy range of injected $\gamma$-rays and the typical counts are about 20 counts s$^{-1}$ for CsI thickness of 2 cm. In orbit, the CZTI and veto detector background is varying uniformly with time with the CZT background count rate in the range of 80 -- 120 count s$^{-1}$ per quadrant and the veto background count range is 300 -- 450 count  s$^{-1}$ per quadrant.

The CZTI can operate in 16 possible modes. Fifteen of these are primary modes, and there is one Secondary Spectral Mode which runs in parallel with other primary modes. But, in normal course of operation data are acquired in only three of them: Normal Mode, Secondary Spectral Mode and SAA mode.  Normal mode is the default mode of operation of the CZTI where complete raw data (144 Mbytes per orbit) are received. Total number of events are recorded in each detector pixel, resulting in a Detector Plane Histogram (DPH). Each count value in the DPH is then divided by the relative quantum efficiency of the corresponding pixel. The array resulting from this is called a Detector Plane Image (DPI). This DPI is a linear combination of shadows of the coded mask cast on the detector plane by sources in the FoV. A library of shadow patterns expected from sources located at different positions on the source plane created by a ray tracing method is used and different methods like Cross correlation, $\chi^2$ shadow fitting, and Richardson-Lucy algorithm (based on Bayesian inference), are applied to get the source position.  The results from the three different methods agree to better than 2 arcmin. The test when repeated at two different energies (60 keV and 122 keV) in the laboratory had shown that at 122 keV, the PSF is seen to widen slightly, by $\sim$10$\%$.  For energies $>$100 keV the mask plate begins to become transparent, reducing shadow contrast and worsening the imaging capability. The results, however, show that decent imaging is still possible at energies as high as 120 keV, with a PSF of less than $\sim$4 arcmin FWHM.

\begin{figure}[h]
\centering
\includegraphics[width=120mm]{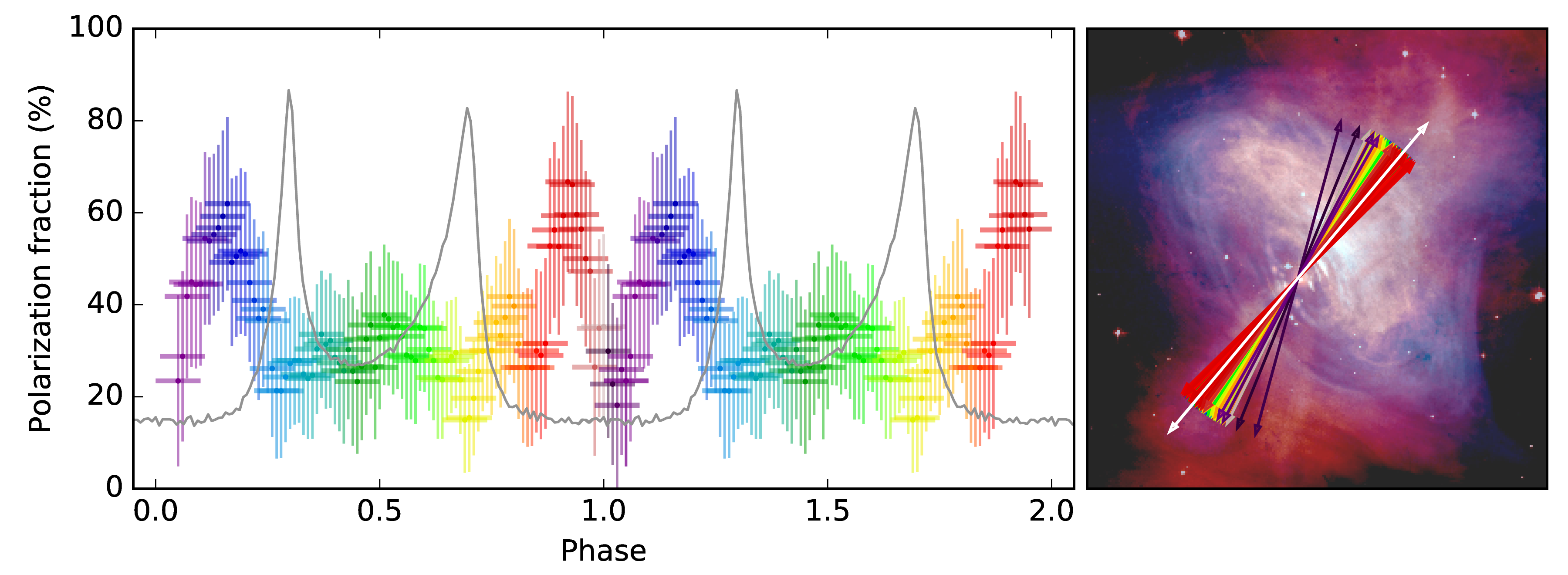}
\caption{Left panel: The grey line shows the brightness of the Crab pulsar as observed by AstroSat CZTI. The horizontal axis (phase) represents time expressed in units of the pulsar’s spin period. Phase 0.0 to 1.0 stands for the full rotation cycle of the pulsar. The same result is shown repeated between phase 1.0 and 2.0, for a clear demonstration of the periodic pattern. Coloured bars indicate the strength of polarized the radiation. Sharp variation of polarization when the brightness is low is a surprising discovery made by AstroSat.  Right panel: Polarization angles shown as arrows in the sky plane plotted over the composite image of the Crab Nebula taken with Chandra and Hubble Observatories with the colour of the arrows representing the phase bin as shown in the plots of the polarization fraction. (Credits: \cite{Vadawale2018}, NASA/CXC/ASU).}
\label{fig:Crab}       
\end{figure}

The pixellated detectors here can be used to map the tracks made by polarized component of X-rays above $>$100 keV.  The CZTI is, therefore, capable of measuring the polarization of bright sources ($>$500 mCrab) in the 100 –- 380 keV region. This capability to measure hard X-ray polarization from bright persistent sources has been demonstrated by observing Crab Nebula (see Fig.~\ref{fig:Crab}; and \cite{Vadawale2018}). The CZTI data obtained confirm the previous results of a strongly polarized off-pulse emission, report a variation in polarization properties within the off-pulse region, and hint at a swing of the polarization angle across the pulse peaks in the Crab pulsar.

\subsection{\textit{CZTI Data Products and Analysis}}

The analysis of data involves image reconstruction by first binning the data to produce DPI, which is then cross correlated with the mask pattern. Significant peaks from this image are picked and a least square fit is then made of the DPI with theoretically computed shadow patterns for sources at these locations. This allows accurate determination of the intensities of these sources, and also helps identify and eliminate spurious peaks that might have been picked from the cross-correlated image. The conversion of L1 data to L2 data involves the following steps:  1) generate the good time interval file using the house keeping information, 2) read the attitude information and generate average position of the satellite for the period of observation, 3) extract the relevant data and generate the raw DPH/DPI, 4) process the detector data (DPH) to generate information about noisy and dead pixels for the time of observation, 5) using information generated in steps 2 and 4, process the raw DPH/DPI and clean them, and generate a detector mask file that records the pixels removed in the cleaning process, 6) perform cross-correlation imaging and pick candidate sources from the image, 7) perform shadow fitting to estimate fluxes of candidate sources, reject insignificant candidates and iterate until all sources have fitted flux values above the detection threshold.  Users of CZTI data are recommended to download L2 data product from astrobrowse website of the ISSDC.  The L2 data  consist of FTOOLS compatible cleaned event file which can be used with any HEASOFT task.  For spectral analysis of a source at the centre of the FOV the default pha file and the corresponding response file available in L2 products can be used in XSPEC.  In case any binning (time,energy) change is required then "cztbindata" can be run on the L2 event file with appropriate inputs. This task as well as other L1 to L2 pipeline elements may be downloaded from ASSC site which also has the corresponding documents and user guides (\url{http://astrosat-ssc.iucaa.in/}).  Studying polarization from a source requires detailed detector simulations in GEANT4 and is currently not accessible to a general user. The interested user may contact the instrument team.

The CZTI has proven to be a very effective All-Sky monitor in the hard ($>$100 keV) X-ray regime, detecting a large number of off-axis transient sources that include $\sim$400 GRBs since launch, and putting highly competitive upper limits on X-ray emissions from gravitational wave sources and fast radio bursts.   However, calculating the source flux or the spectrum requires knowledge of the direction and energy dependent attenuation of the radiation incident upon the detector.  A GEANT4-based mass model of CZTI and AstroSat that can be used to simulate the satellite response to the incident radiation, and to calculate an effective “response file” for converting the source counts into fluxes and spectra has been developed by \cite{Mate2021}. This has been used to develop a program called CIFT: the CZTI Interface for Fast Transients (\cite{Sharma2021}).  The Earth Occultation Technique (EOT) has also been developed to study persistent sources at large off-axis angles and demonstrated by measuring the flux of the Crab nebula and pulsar, and measuring the variability of Cyg X-1 (\cite{Singhal2021}).  Measurements of polarization of prompt $\gamma$-ray emission from GRBs has also been reported by \cite{Chattopadhyay2019}.

For more details about the CZTI instrument, please see \cite{Bhalerao2017}.

\section{\textit{Scanning Sky Monitor (SSM)}}

\begin{figure}[h]
\centering
\includegraphics[width=0.54\textwidth,clip]{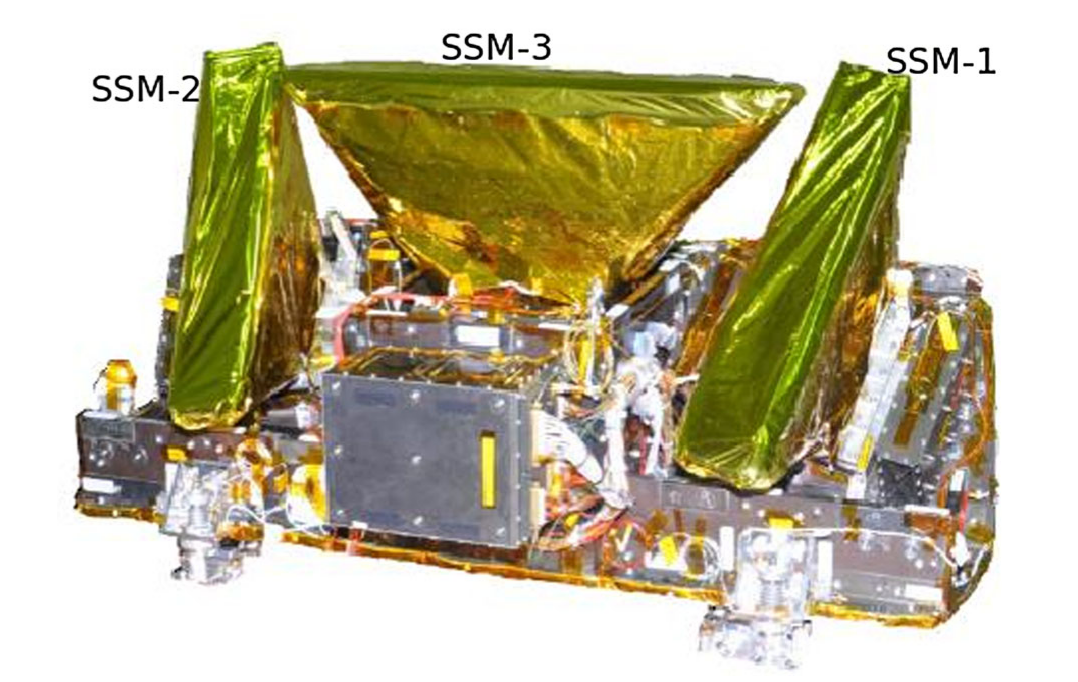}
\includegraphics[width=0.45\textwidth,clip]{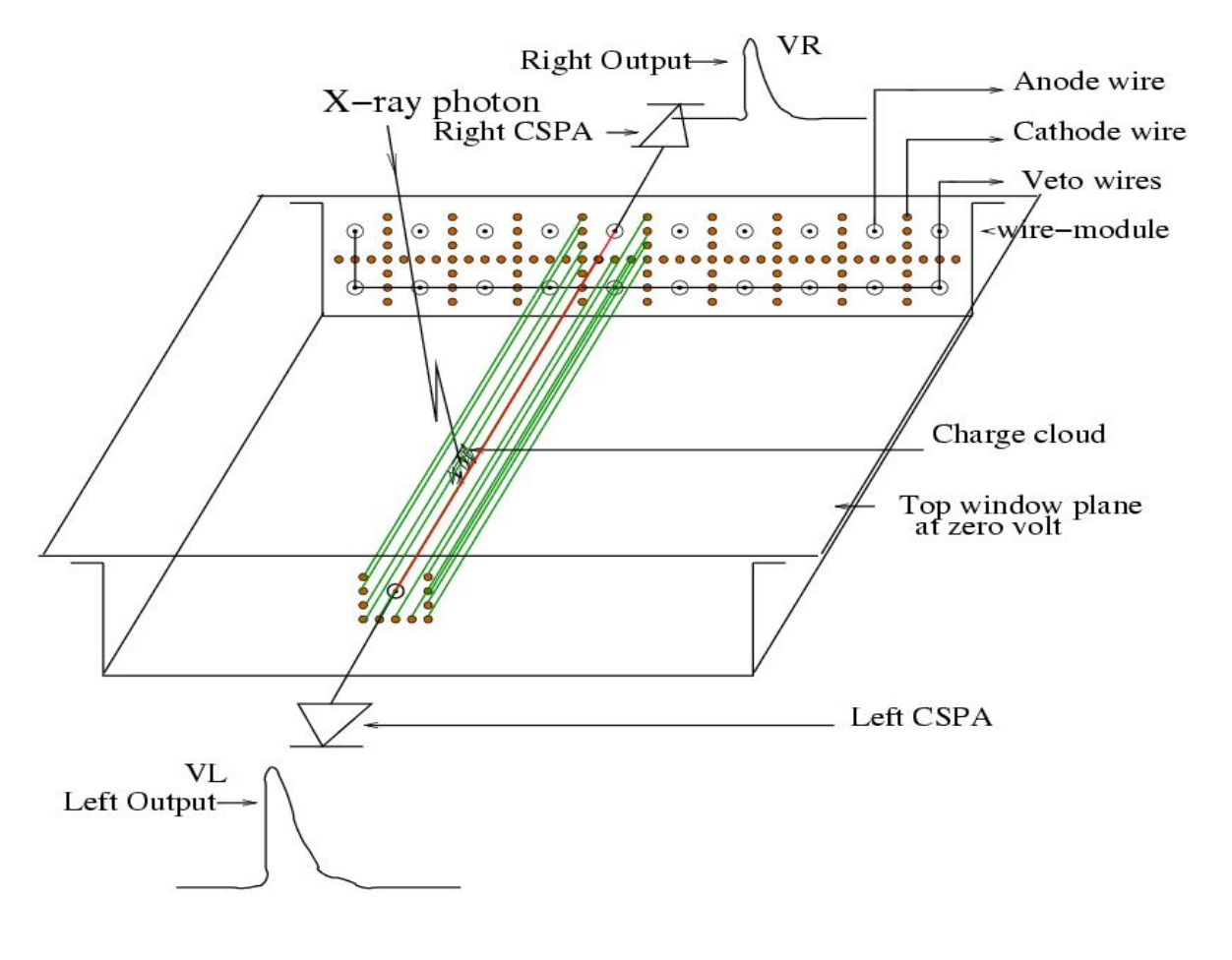}
\includegraphics[width=0.45\textwidth,clip]{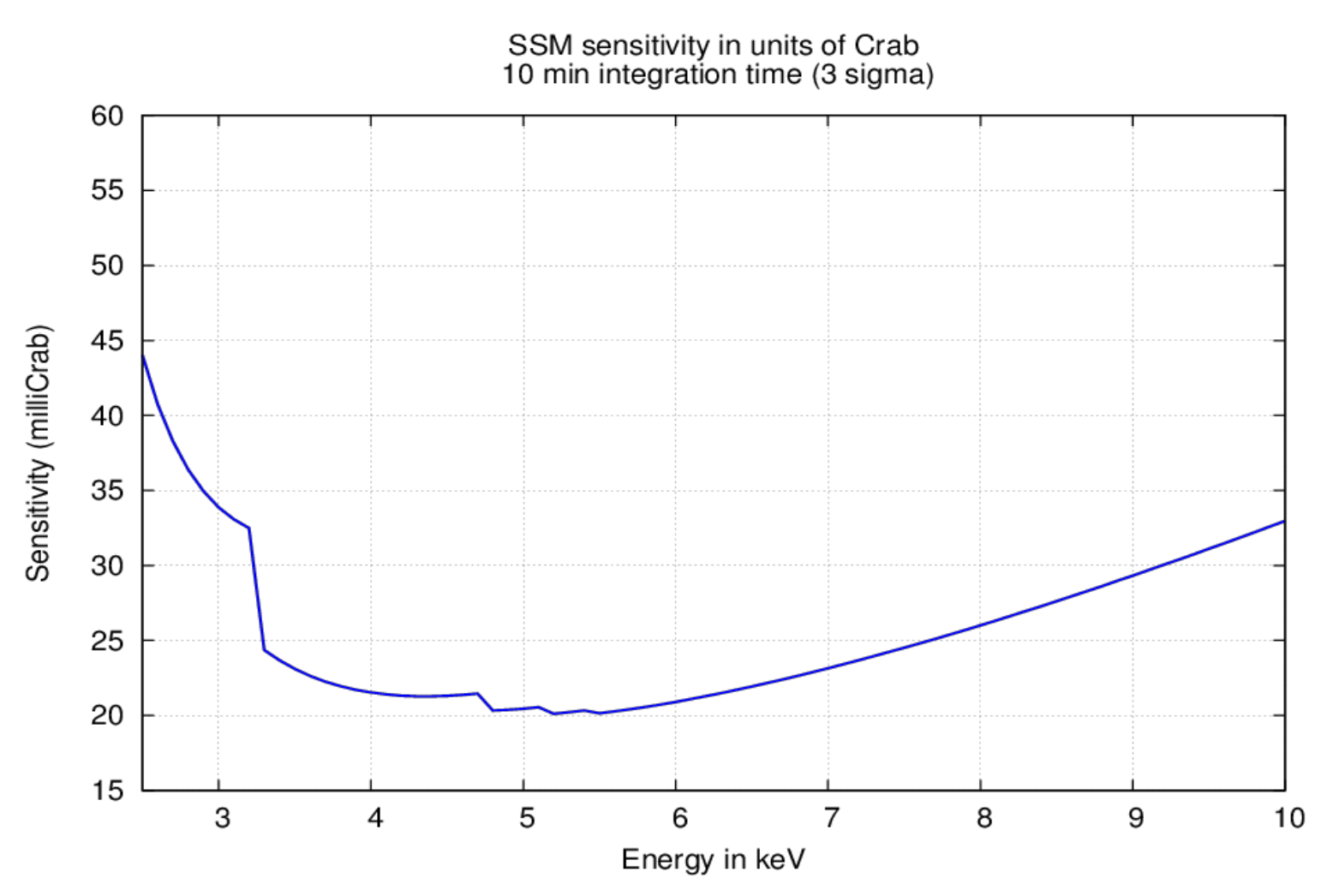}
\caption{(Top-left) All three units of the SSM mounted on a single rotatable platform. (Top right)  A schematic view of the cells inside an SSM detector. (Bottom panel) The detection sensitivity of the SSM as a function of energy of the source.  Credits: \cite{Ramadevi2017a}, \cite{Ramadevi2017b}, \cite{Ramadevi2018}.}
\label{fig:SSM}       
\end{figure}

The SSM is designed to scan the sky for known bright X-ray sources and unknown X-ray sources in the 2.5 - 10 keV energy range.   It  weighs 64.7 kg and has an envelope of 1.200 m (length) x 0.563 m (width) x 0.543 m (height). The SSM is made of three nearly identical units of position sensitive gas-filled proportional counters each having a coded-mask. A coded mask pattern with 50$\%$ transparency joined sideways, provides a spatial resolution of $\sim$1 mm at 6 keV (angular resolution $\sim$12 arcmin on the sky) in one direction.  In a direction perpendicular to the coding direction it is 2.5 degree.  Each SSM unit scans the sky in one dimension over a FoV of 22 x 100 degree (for SSMs on the edges) and $\sim$26.8 x 100 degree (for the central SSM).  All three units of SSM mounted on a single platform are shown in Fig.~\ref{fig:SSM} (top-left). Each SSM with its associated electronics is mounted on a rotating platform to scan the sky on the anti-sun side of the spacecraft whereas the processing electronics is mounted inside the satellite.  Each unit has eight anodes with a high voltage of $\sim$1500 volts. Each SSM contains xenon (25$\%$), argon and methane (75$\%$ P-10) gas mixture at 800 torr held by window of 50$\mu$ thick aluminised Mylar.  Fig.~\ref{fig:SSM} (top-right) also shows a schematic view of the cells inside an SSM detector. The top layer consists of position-sensitive anode wires surrounded by wire-walled cathode forming individual cells of size 12 cm$^2$, and 6 cm long. Charge is collected at the anode for every photon that is incident and converted to a voltage pulse using a charge sensitive pre-amplifier (CSPA) on either side of the anode. There are seventeen such chains for each SSM detector unit, sixteen of them are connected to both ends of eight anodes and one of them is connected to a veto layer used to reject charge particles using anti-coincidence when (a) signals are present in both the top anode layer as well as the bottom veto layer simultaneously, or (b) more than one anode has an event simultaneously.  The processing electronics (PE) is an FPGA-based unit which stores the accepted  events in a buffer memory and interfaces with the Data Handling system of the spacecraft. The PE also acts as both telecommand and telemetry interface for the complete SSM electronics system. The time of incidence of every photon recorded as an event is tagged by the PE using an onboard SSM clock. This clock is periodically synchronised with the spacecraft BMU clock.  The effective area of SSM is 53 cm$^2$ at 5 keV (11 cm$^2$ at 2.5 keV). The energy resolution is 25$\%$ at 6 keV and its sensitivity is ~$\sim$8 mCrab for 600 s exposure time at 3$\sigma$ confidence.  The sensitivity of the SSM as a function of energy is shown in Fig.~\ref{fig:SSM} (bottom panel).  

The SSM produces two types of data: temporal and time-tagged. Temporal data give the counts detected every 100 ms while the Time-tagged data give the information of every event detected in the detector.  The time-tagged data provide energy of the incident photon and the position of incidence can also be calculated.  Using this data the shadow pattern cast by different sources in the fov of the SSM is derived, and later used to derive the position and flux of the sources in the FoV.  The SSM2 unit, one of the edge cameras has been switched OFF since August 2016, as it developed gas leak by the middle of February 2016.  Two other units: SSM1 (one of the edge cameras) and SSM3 (the central camera) are continuing to observe the X-ray sky in a step-and-stare mode of operation.  The SSM1 unit has, however, shown a reduced efficiency and a correction factor is applied to the intensity of the sources observed by this unit. As of November 2021, the gain in the SSM1 has increased beyond tolerable limits and it is likely to be shut down soon.  X-ray light curves are generated routinely and made available to the public. The URL for accessing SSM Web is \url{https://webapps.issdc.gov.in/SSM_Web/index.jsp} and it can also be accessed through ISSDC website \url{https://issdc.gov.in --> Astrosat --> SSM Web}.  The software pipeline developed for reducing the raw SSM data from the Indian Space Science Data Centre (ISSDC) to generate the L1 and L2 and to send it back to ISSDC for dissemination has been developed and is described by \cite{Ravishankar2021}.

For more details about SSM please see \cite{Ramadevi2017a} and \cite{Ramadevi2017b}.  Results from the onboard calibration of SSM are described in \cite{Sarwada2021}.  For the performance of the SSM and the results obtained from the observations of Transient X-ray sources the reader is referred to \cite{Ramadevi2018}.

\section{\textit{Charged Particle Monitor (CPM)}}

\begin{figure}[h]
\centering
\includegraphics[width=0.45\textwidth,clip]{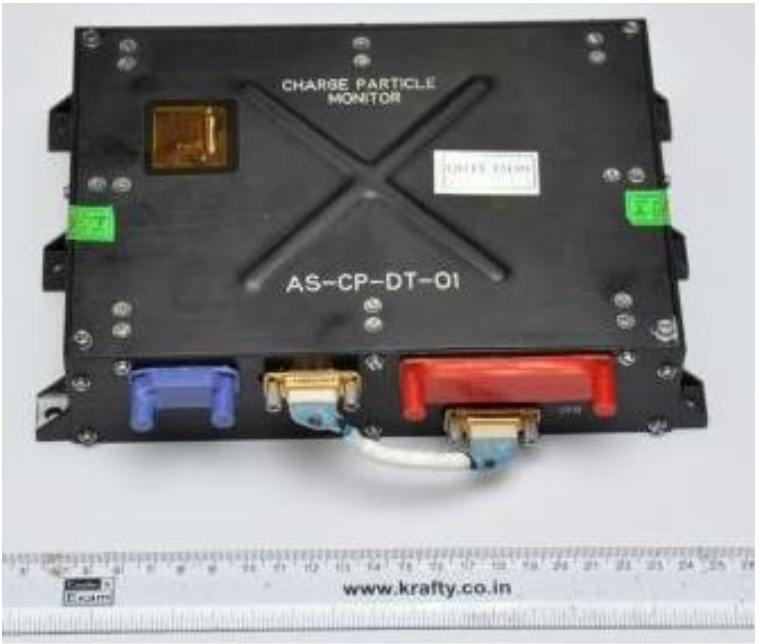}
\includegraphics[width=0.48\textwidth,clip]{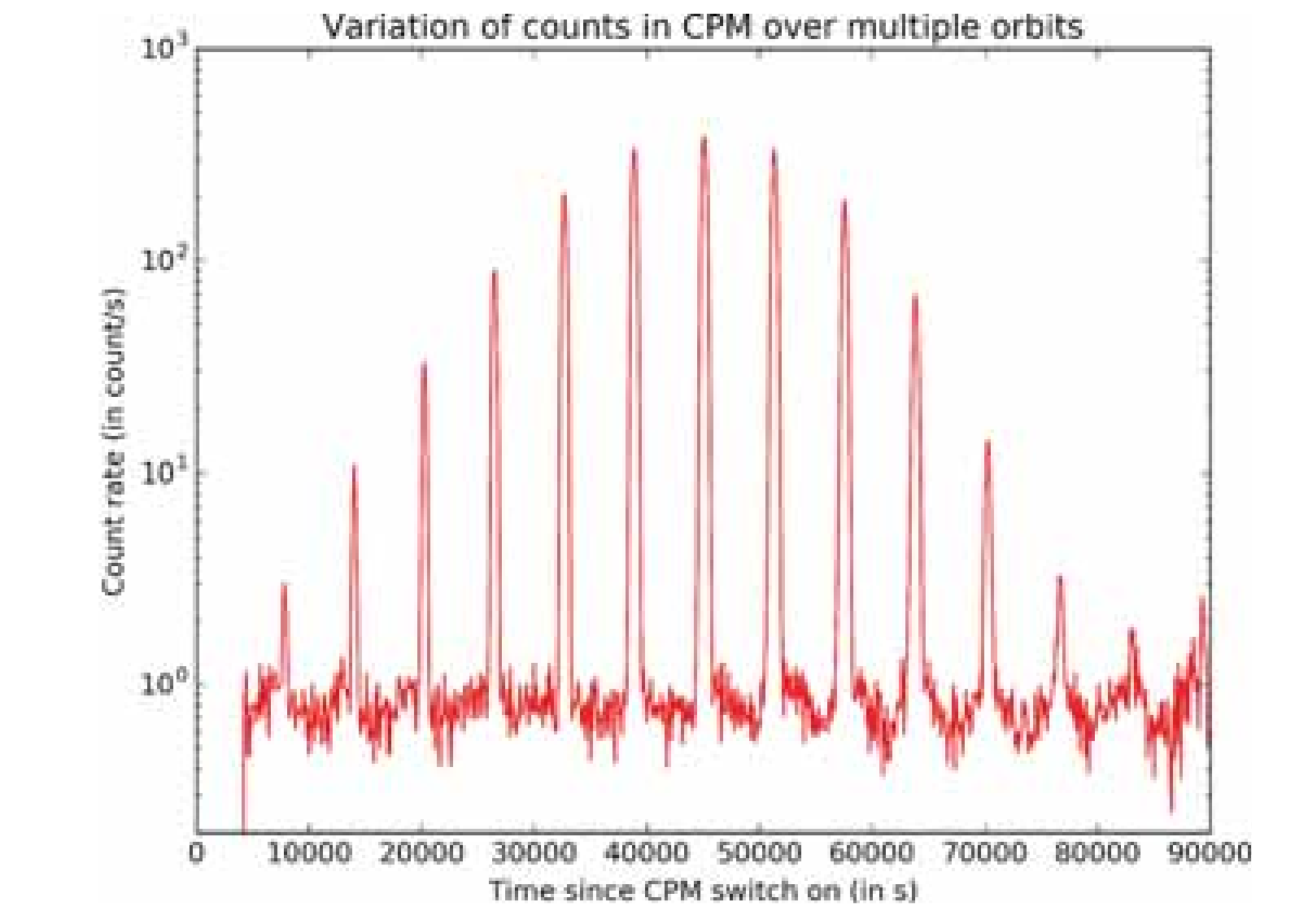}
\includegraphics[width=0.49\textwidth,clip]{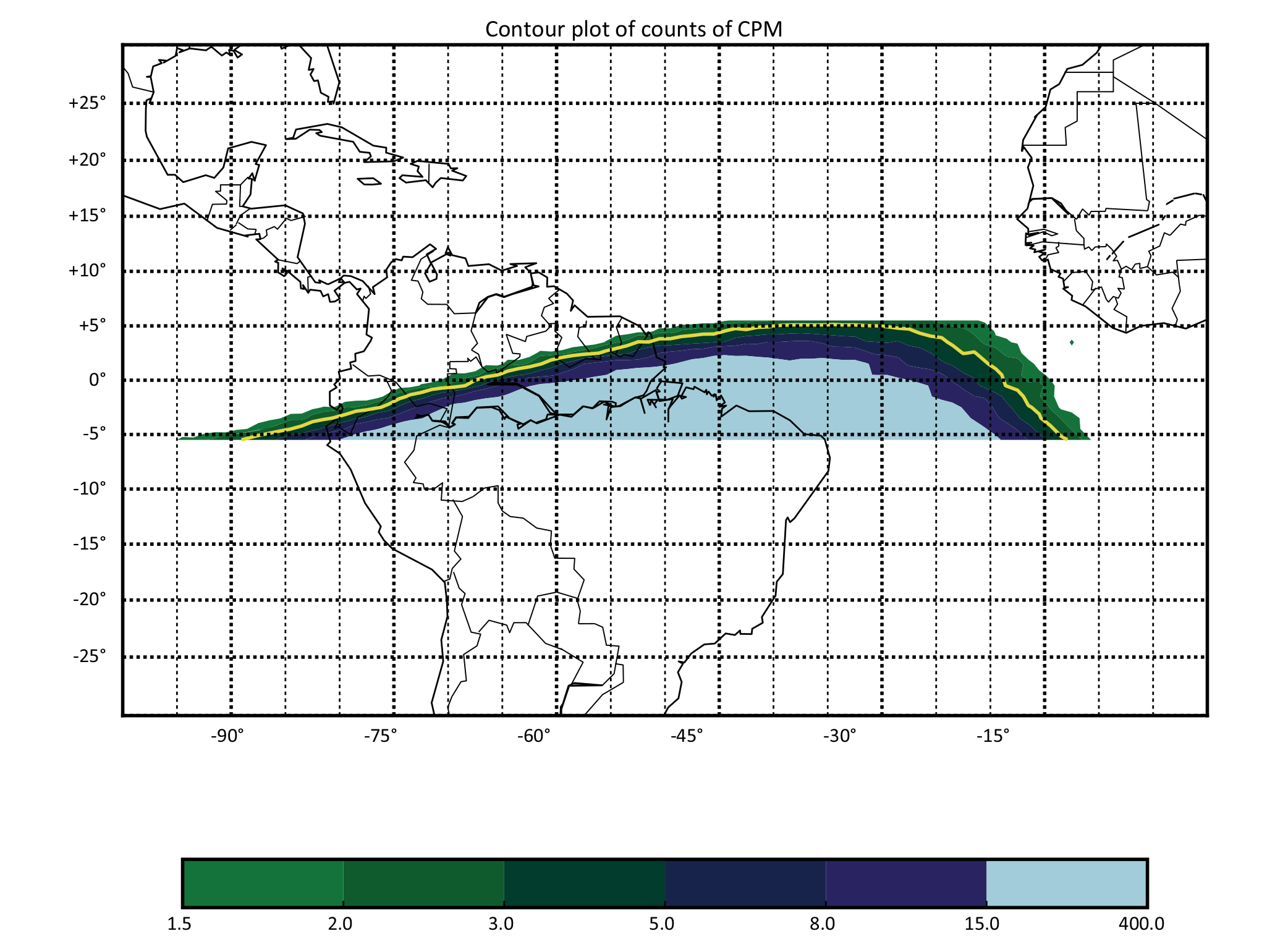}
\caption{(Left) The flight unit of the CPM. (Top-right) Count rates as observed by the CPM. The recurring peaks denote the sections of SAA region traversed in an orbit. The highest peak is when the satellite is the deepest into the SAA region. Variations in peak height reflect the shift of the region of the SAA sampled by successive orbits of the satellite. (Bottom) Contour maps count rates from the SAA region based on one month’s accumulated data. The count rates vary from $\sim$1 in non-SAA region to  a peak value of $\sim$400 count s$^{-1}$. The average count rate in the SAA region is $\sim$280 s$^{-1}$. The boundary value used for the SAA region is 3 count s$^{-1}$ and is shown as a yellow line in the figure. Credits: \cite{Rao2017}.}
\label{fig:CPM}       
\end{figure}

The CPM is an auxiliary scientific payload that is used to warn other scientific instruments about the passage through high particle background regions such as the South Atlantic Anomaly.  It has a size of 222 mm x 138 mm  48 mm and it weighs 2 kg.  The CPM consists of a Cesium Iodide Thallium activated (CsI(Tl)) scintillation as a detector with an open area of 1 cm$^2$.  A 10 mm cube of CsI(TI) crystal having teflon reflective material is coupled to a window of same area of a Si-PIN photodiode that has a good response to the visible spectrum (average efficiency of $\sim$50$\%$),  high sensitivity, low dark current and good energy resolution. The top side of the detector is covered with a very thin sheet (25.4 $\mu$m) of Kapton which along with a protective 10 $\mu$m teflon tape gives an effective low-energy threshold of about 1.2 MeV and provides  sufficient protection. The electronic threshold is, however, programmable and kept at 0.5 MeV. A thin Copper box also acts as a radiation shield.  Ionising radiations that create photons in the scintillating crystal are captured by a Si-Pin diode placed at the bottom surface of the CsI(Tl) crystal.  The light output peaks at $\sim$580 nm with its intensity being proportional to the energy of the incident radiation. This light is proportionally converted to an electrical signal by the photodiode. A charge sensitive preamplifier (CSPA) is used to amplify the photodiode output to milli-volt range.  The detector module is mounted on an electronics card along with other electronics like amplifiers, discriminators, and an FPGA for handling interfaces and detector logic. A Power Module is mounted below this card to generate the required voltages from the Satellite Raw Bus of +42 Volts. The detector module generates a pulse corresponding to every incident energetic particle and this pulse is passed through a post amplifier, where milli-volt signal is converted to volt level. The total gain of amplifier is kept around 500 with sufficient bandwidth such that it generates 0.5 $\mu$s rise time pulses. These amplified pulses are passed through a Low Level Discriminator (LLD) circuit to cut-off noise and select the charge particle signals above a defined value of about 0.5 MeV. The LLD reference voltage value is programmable from ground station through telecommand. It is a 12-bit DAC and 4096 discrete values are possible. The pulse signal from LLD circuit output is digitized and fed to a 14-bit counter. The 14-bit counter is a free running counter gated every 5 sec (default). Gate time is also programmable from ground through telecommand. All digital processing electronics is embedded in the FPGA. The telemetry and telecommand interface with the satellite bus is also embedded in the FPGA.  An image of the CPM is shown in Fig.~\ref{fig:CPM} (top-left panel). 

The performance of the CPM is shown in the top-right and bottom panels of Fig.~\ref{fig:CPM}, showing its utility for the screening of the background regions for the SXT, LAXPC and CZTI instruments. For more details of the CPM, please see \cite{Rao2017}.

\section{\textit{Conclusions}}

A brief overview of the AstroSat mission, its scientific payload including their description, characteristics, onboard performance, software and pipelines for reducing the observational data that are available publicly, has been presented here.

\section{\textit{Acknowledgements}}

I acknowledge the help and support of many individuals named in the credits of the figures who have allowed me to use the figures from their publications, the scientists and engineers running the Payload Operation Centres for all the instruments at IIA, TIFR and IUCAA, the staff at ISSDC for maintaining and running the AstroSat, the AstroSat Science Support Cell at IUCAA, the ISRO HQ and the Principal Investigators of AstroSat.

\end{document}